\documentclass[fleqn,usenatbib]{mnras}

\usepackage{newtxtext,newtxmath}

\usepackage[T1]{fontenc}
\usepackage{ae,aecompl}

\usepackage{graphicx}	
\usepackage{amsmath}	
\usepackage{amssymb}	
\usepackage[colorinlistoftodos,prependcaption]{todonotes} 
\usepackage{siunitx} 

\newcommand\eqn[1]{equation~\ref{#1}}

\newcommand\fig[1]{Figure~\ref{#1}}

\newcommand\sect[1]{Section~\ref{#1}}

\newcommand{\dx}[1]{\mathrm{d}{#1}\,}

\DeclareSIUnit \h {\mbox{$h$}}
\DeclareSIUnit \hinv {\mbox{$h^{-1}$}}
\DeclareSIUnit \deg {\mbox{deg}}
\DeclareSIUnit \msun {\mbox{$M_{\odot}$}}

\DeclareSIUnit \parsec {pc}
\DeclareSIUnit \kilaparsec {kpc}
\DeclareSIUnit \megaparsec {Mpc}
\DeclareSIUnit \gigaparsec {Gpc}
\DeclareSIUnit \arcminutes {arcminutes}

\newcommand{\om}{\ensuremath{\Omega_{\mathrm m}}}

\newcommand{\sig}{\ensuremath{\sigma_8}}
\newcommand{\lcdm}{$\Lambda$CDM}

\newcommand{\gammat}{\ensuremath{\gamma_{t}(\theta)}}

\newcommand{\dsigr}{\ensuremath{\Delta\Sigma(R)}}
\newcommand{\dsigobsr}{\ensuremath{\Delta\Sigma^{\mathrm{obs}}(R)}}
\newcommand{\dsigmodr}{\ensuremath{\Delta\Sigma^{\mathrm{model}}(R)}}

\newcommand{\sigcrit}{\ensuremath{\Sigma_{\mathrm{crit}}}}
\newcommand{\rmin}{\ensuremath{r_{\mathrm{min}}}}
\newcommand{\thetamin}{\ensuremath{\theta_{\mathrm{min}}}}
\newcommand{\zlens}{\ensuremath{z_{\mathrm{l}}}}
\newcommand{\zsource}{\ensuremath{z_{\mathrm{s}}}}
\newcommand{\nzlens}{\ensuremath{n_{\mathrm{l}}}(z_{\mathrm{l}})}
\newcommand{\nzsource}{\ensuremath{n_{\mathrm{s}}}(z_{\mathrm{s}})}
\newcommand{\nzlensi}{\ensuremath{n_{\mathrm{l},i}}(z_{\mathrm{l}})}
\newcommand{\nzsourcej}{\ensuremath{n_{\mathrm{s},j}}(z_{\mathrm{s}})}
\newcommand{\xigm}{\ensuremath{\xi_{\mathrm{gm}}}}
\newcommand{\xigg}{\ensuremath{\xi_{\mathrm{gg}}}}
\newcommand{\ximm}{\ensuremath{\xi_{\mathrm{mm}}}}

\newcommand{\nlens}{\ensuremath{n_{\mathrm{lens}}}}
\newcommand{\nsrc}{\ensuremath{n_{\mathrm{src}}}}

\DeclareMathOperator{\matc}{\mathbfss{C}}

\DeclareMathOperator{\matn}{\mathbfss{N}}

\title[Small-scale information in galaxy-galaxy lensing]{Controlling and leveraging small-scale information in tomographic galaxy-galaxy lensing}

\author[N. MacCrann et al.]{
Niall MacCrann$^{1,2}$\thanks{E-mail: maccrann.2@osu.edu},
Jonathan Blazek$^{3}$,
Bhuvnesh Jain$^{4}$
and Elisabeth Krause$^{5}$
\\
$^{1}$ Center for Cosmology and Astro-Particle Physics, The Ohio State University, Columbus, OH 43210, USA\\
$^{2}$ Department of Physics, The Ohio State University, Columbus, OH 43210, USA\\
$^{3}$ Institute of Physics, Laboratory of Astrophysics, \'Ecole Polytechnique F\'ed\'erale de Lausanne (EPFL),\\ Observatoire de Sauverny, 1290 Versoix, Switzerland\\
$^{4}$ Department of Physics and Astronomy, University of Pennsylvania, Philadelphia, PA 19104, USA\\
$^{5}$ Department of Astronomy/Steward Observatory, 933 North Cherry Avenue, Tucson, AZ 85721-0065, USA\\
}

\date{Accepted XXX. Received YYY; in original form ZZZ}

\pubyear{2015}

\begin{document}
\label{firstpage}
\pagerange{\pageref{firstpage}--\pageref{lastpage}}
\maketitle

\begin{abstract}
The tangential shear signal receives contributions from physical scales in the galaxy-matter correlation function well below the transverse scale at which it is measured. Since small scales are difficult to model, this non-locality has generally required stringent scale cuts  or new statistics for cosmological analyses. Using the fact that uncertainty in these contributions corresponds to an uncertainty in the enclosed projected mass around the lens, we  provide an analytic marginalization  scheme  to  account  for  this. Our approach enables  the  inclusion  of measurements on smaller scales without requiring numerical sampling over extra free parameters.  We extend the analytic marginalization formalism to retain cosmographic (``shear-ratio") information from small-scale measurements that would otherwise be removed due to modeling uncertainties, again without requiring the addition of extra sampling parameters. We test the methodology using simulated likelihood analysis of a DES Year 5-like galaxy-galaxy lensing and galaxy clustering datavector. We demonstrate that we can remove parameter biases due to the presence of an un-modeled 1-halo contamination of the galaxy-galaxy lensing signal, and use the shear-ratio information on small scales to improve cosmological parameter constraints.
\end{abstract}


\section{Introduction}

\vspace{0cm}

The observed shapes of distant galaxies are distorted due to variations in the gravitational potential experienced by emitted light on its path to the observer, an effect known as gravitational lensing. In the weak lensing regime, a small change in the observed ellipticity of such a \emph{source} galaxy is generated, known as a shear. Coherent structure in the intervening density field generates coherent patterns in the observed shear field. For example, a net tangential alignment or \emph{tangential shear} of source galaxies is produced around overdense regions in the intervening density field. 

Since galaxies also trace overdense regions, we can measure the average tangential shear of source galaxies around these tracers, also known as \emph{lens} galaxies, to probe the relationship between lens galaxy and matter densities. This sort of measurement is known as \emph{galaxy-galaxy lensing}, and since early detections by \citet{tyson84} and \citet{brainerd96}, it has been measured at increasing signal-to-noise (e.g. \citealt{choi2012,cacciato13,mandelbaum13,velander14,clampitt17,pratsanchez18}), and precision measurements from  state-of-the-art photometric imaging surveys have been used for cosmological parameter estimation \citep{keypaper,vanuitert18,joudaki18,singh18}.

The galaxy-galaxy lensing signal depends on the total matter distribution around the lens galaxies, or the galaxy-matter cross-correlation function $\xigm(r)$. In the halo-model picture \citep{seljak00,peacock00}, on small scales the measurement is most sensitive to the properties of the halos populated by the lens galaxies, for example the mean halo mass. Hence galaxy-galaxy lensing has been used to characterize the relation between halo mass and baryonic content of galaxies (e.g. \citealt{leauthaud12, viola15, vanuitert16}). On larger scales, galaxy-galaxy lensing has been combined with galaxy clustering to simultaneously constrain the galaxy bias, and cosmological parameters, in particular the matter density and matter clustering amplitude at low redshift. Especially when combined with external constraints from e.g. the cosmic microwave background, this combination can also provide competitive constraints on the dark energy equation of state \citep{weinberg13}. These constraints will only improve with upcoming stage IV surveys such as the Large Synoptic Survey Telescope\footnote{\url{http://www.lsst.org}} (LSST), Euclid\footnote{\url{http://sci.esa.int/science-e/www/area/index.cfm?fareaid=102}} and the Wide-Field Infrared Survey Telescope\footnote{\url{http://wfirst.gsfc.nasa.gov}} (WFIRST), which will dramatically increase the volume of high quality weak-lensing data available.

There are typically significant observational and theoretical challenges in performing a galaxy-galaxy lensing analysis \citep{mandelbaum18}. In the former category, the shear must be estimated with high accuracy from images of faint source galaxies that are typically noisy and blurred by a point spread function, an ongoing challenge in the weak lensing community (e.g. \citealt{great08handbook,great10,great3handbook}). Furthermore, interpreting the signal requires redshift information for both lens and source galaxies, which generally requires estimating photometric redshifts from noisy flux estimates in a small number (typically around 5) of  optical or near-infrared bands (e.g. \citealt{hildebrandt17,y1photoz,tanaka18}). 

There are also significant theoretical challenges when attempting to model the galaxy-galaxy lensing signal, which generally becomes more difficult at smaller scales. In order to predict the signal, an accurate prediction for the galaxy-matter correlation function $\xigm(r)$ is required for some range of physical scales $r$. On sufficiently large scales we expect linear bias to hold (e.g. \citealt{fry93,kaiser84}), such that $\xigm(r)=b\ximm(r)$, where $b$ is the linear galaxy bias, an unknown constant that can be marginalized over and $\ximm(r)$ is the matter correlation function. A higher-order perturbative modelling approach may be successful in predicting $\xigm(r)$ at smaller, mildly nonlinear scales (see \citealt{desjacques18} for a recent review). A perturbative approach will likely fail on scales approaching the 1-halo regime, but here a model which assumes some halo occupation  distribution \citep{peacock00,seljak00,berlind02,wechsler18} combined with an accurate prediction for the clustering of dark matter halos may be successful (e.g. \citealt{cacciato13,nishmichi18,wibking19}). Even this approach will break down on galactic scales where galactic astrophysics will affect the matter distribution in and around the lens galaxy. 

The important point is that whatever modeling approach is taken, it is crucial to ensure that the measurement is only sensitive to scales in $\xigm(r)$ where that modelling approach is sufficiently accurate. In \sect{sec:theory} we describe how the galaxy-galaxy signal receives a non-local contribution that depends on scales in $\xigm(r)$ that are much smaller than the separation at which the measurement is made (i.e. the impact parameter in the lens plane). It was this potential non-local contribution that motivated the use of a larger minimum scale for galaxy-galaxy lensing than for galaxy clustering in the combined clustering, galaxy-galaxy lensing and cosmic shear analysis of Dark Energy Survey (DES) Year 1 data in \citet{methodpaper,keypaper}.
We demonstrate how this non-local contribution can be accounted for in parameter estimation, and use analytic marginalization \citep{bridle02} to avoid adding extra sampling parameters. 

When galaxy-galaxy lensing of a given lens sample is measured from multiple sources redshifts, some limited information can be extracted even in the absence of a model for the galaxy-matter correlation function. This is often referred to as \emph{shear-ratio} information; since the ratio of the signals measured from two different source redshifts depends only on the Universe's geometry \citep{jain03,hu04,bernstein04}. In \sect{sec:smallscalemarg} we extend the aforementioned analytic marginalization formalism to allow the retention of shear-ratio information from small-scale galaxy-galaxy lensing measurements that would otherwise be excluded due to modelling uncertainties. 
Our use of analytic marginalization for both these problems makes our methods much more useful for cosmological parameter estimation from weak lensing surveys; without this the extra tens or hundreds of sampling parameters may lead to significant increases in convergence time for MCMC-based inference. We demonstrate the utility of our methodology by simulating cosmological parameter inference from a DES Year 5-like galaxy-galaxy lensing and galaxy clustering datavector in \sect{sec:cosmotests}. 

We conclude and discuss some potential limitations of the methodology in \sect{sec:discussion}.

\section{The point-mass contribution to tangential shear }\label{sec:pm}

We start in \sect{sec:theory} by describing how physical scales in $\xigm$ contribute to the galaxy-galaxy lensing signal, and how the non-local contribution from small physical scales can be marginalized over. We draw in particular on \citet{baldauf10} (also see a recent treatment in \citealt{singh18}). We discuss the use of analytic marginalization in \sect{sec:anmarg}, and compare to the approach of \citet{baldauf10} in \sect{sec:upsilon}. We extend the formalism to a tomographic tangential shear measurement in \sect{sec:tomo}, and demonstrate the effectiveness of our approach in recovering unbiased parameters in \sect{sec:simpletests}.

\subsection{Theory}\label{sec:theory}
A lens galaxy sample at angular diameter distance $D_l$ generates a mean tangential shear, $\gammat$ (e.g. \citealt{hu04}) 
\begin{equation}
\gamma_t(\theta = R/D_{\mathrm{l}}) = \frac{\dsigr}{\sigcrit}
\end{equation}
where 
\begin{equation}
\dsigr = \overline{\Sigma}(0,R) - \Sigma(R) \label{eq:deltasig}
\end{equation}
and $\Sigma(R)$ is the excess mean surface mass density at transverse physical separation $R$ from the lens, given by the projection of the three-dimensional galaxy-matter correlation function $\xigm$ over line-of-sight distance $\Pi$:
\begin{equation}
\Sigma(R) = \overline{\rho}_m \int_{-\infty}^{\infty} \dx{\Pi} \left[1 + \xi_{gm}\left( \sqrt{R^2+\Pi^2}\right) \right],
\end{equation}
where $\overline{\rho}_m$ is the mean matter density.
$\overline{\Sigma}(R_1,R_2)$ is the mean surface mass density averaged between $R_1$ and $R_2$
\begin{equation}
\overline{\Sigma}(R_1,R_2) = \frac{2}{R_2^2-R_1^2} \int_{R_1}^{R_2} R' dR' \Sigma(R').
\end{equation}
\sigcrit\ is a geometrical factor that determines how the amplitude of the signal depends on lens and source redshift, and for a single source redshift plane at angular diameter distance $D_s$, is given by
\begin{equation}
    \sigcrit^{-1} = \begin{cases}
    (1+\zlens)\frac{4\pi G}{c^2}\frac{D_l( D_s-D_l)}{D_s} & \text{if $D_s>D_l$}\\
    0 & \text{otherwise}
    \end{cases}
\end{equation}\label{eq:sigcrit}
where $z_l$ is the lens redshift.

The presence of $\overline{\Sigma}(0,R)$ in \eqn{eq:deltasig}
makes clear the non-locality of $\dsigr$ (and therefore $\gammat$);  
this term depends on the distribution of mass around the lens on all scales up to $R$, or equivalently, on $\xigm(r)$ for all $0<r<R$. As a consequence, $\dsigr$ and $\gammat$ can be sensitive to the mass distribution 
on one-halo scales, where a perturbative modeling approach will break down, even when measured at separations $R$ that correspond to much larger physical scales in the lens plane.
There is extensive discussion of this effect in \citet{baldauf10} who propose an estimator-based approach for dealing with this non-locality that we discuss in \sect{sec:upsilon}. We note here that a projected galaxy clustering measurement, $w_{\mathrm{gg}}(R)$ does not suffer from this effect - here the minimum physical scale probed in the three-dimensional correlation function $\xigg(r)$  is the same as the transverse separation $R$.


If we assume we can model the galaxy-matter correlation function $\xigm(r)$ only down to some minimum scale $\rmin$, we can account for the contribution from scales below $\rmin$ in the following way. For $R>\rmin$ we can decompose $\overline{\Sigma}(0,R)$ into two terms 
\begin{equation}
\overline{\Sigma}(0,R) = \frac{\rmin^2 \overline{\Sigma}(0,\rmin)}{R^2} + \frac{(R^2-\rmin^2) \overline{\Sigma}(\rmin, R)}{R^2}.
\label{eq:sigrdecomp}
\end{equation}

Only the first term in \eqn{eq:sigrdecomp} is beyond our ability to model accurately (the second term requires only $\xigm$ at $r>\rmin$). This first term has $1/R^2$ scale dependence, so for $R>\rmin$, any bias in our model due to inaccurate prediction of $\xigm(r<\rmin)$ has a simple $1/R^2$ scale dependence. 

Hence, for $R>\rmin$, we can model \dsigr\  as 
\begin{equation}
\dsigr = \Delta\Sigma^{\mathrm{model}}(R) + B/R^2 \label{eq:pointmass}
\end{equation}
where $\Delta\Sigma^{\mathrm{model}}(R)$ is the prediction based on a model for $\xigm(r)$ that is correct for scales $r>\rmin$, but can be arbitrarily wrong for $r<\rmin$, and $B$ is some unknown constant that we can marginalize over. 

Note that the first term in \eqn{eq:sigrdecomp} is just the tangential shear contribution from the excess mass enclosed in a cylinder of radius $\rmin$. For transverse scales $R$ larger than $\rmin$ this has the same lensing signal as a point-mass located at $R=0$, hence in the following we will refer to this contribution as the \emph{point-mass} contribution. However, the constant $B$ in \eqn{eq:pointmass} does not correspond exactly to this enclosed mass if our model for $\xigm(r)$ makes a non-zero prediction for $\overline{\Sigma}(0,\rmin)$. In this case 
\begin{equation}
B=\frac{\delta M}{\pi R^2 \sigcrit} 
\end{equation}
where $\delta M$ is the bias in the model prediction for the enclosed mass i.e. this term
accounts for inaccuracies in the enclosed mass prediction. We note that for a given lens galaxy sample, $B$ will be a function of lens redshift as well as $R$ i.e. $B=B(\zlens, R)$.

We can also write $B$ in terms of systematic bias on the galaxy-matter correlation function prediction, $\xigm^{\mathrm{bias}}(r) \equiv \xigm^{\mathrm{model}}(r)-\xigm^{\mathrm{true}}(r)$ i.e. the difference between our model for $\xigm(r)$ and the truth,
\begin{equation}
    B = \frac{2}{R^2 \sigcrit}\int_{-\infty}^{\infty} \dx{\Pi} \int_{0}^{R} R'\dx{R'}\left[1+\xigm^{\mathrm{bias}}\left(\sqrt{R^2+\Pi^2}\right)\right].
\end{equation}
A prior on $B$ could then be constructed from a scale-dependent prior on $\xigm^{\mathrm{bias}}(r)$ (see e.g. \citealt{baldauf16} for more discussion of the inclusion of such theoretical uncertainties in cosmological parameter esitmation).

\begin{figure}
\includegraphics[width=\columnwidth]{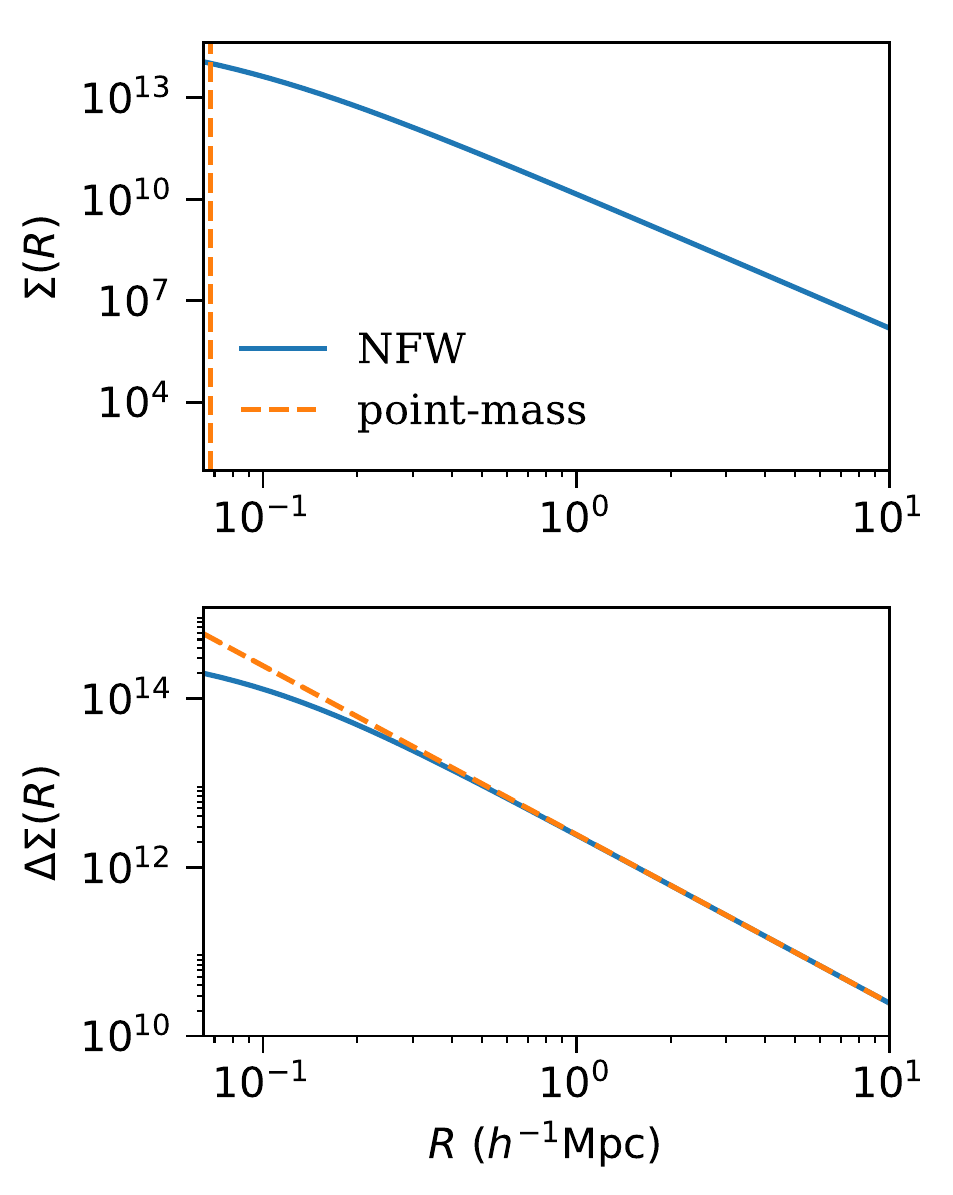}
\caption[]{The surface mass density $\Sigma(R)$ (top panel) and surface mass density contrast $\Delta\Sigma(R)$ (bottom panel) for a smoothly truncated NFW halo (blue solid lines, see \sect{sec:simpletests} for details) and a point-mass of the same mass (orange dashed lines). While the two cases have very different $\Sigma(R)$, they have very similar $\Delta\Sigma(R)$ except at very small scales. Note that in the point-mass case $\Sigma(R)$ is non-zero only at $R=0$; this is represented in the upper-panel by a vertical line at the left edge of the plot.}
\label{fig:nfw}
\end{figure}

In \sect{sec:simpletests}, we perform tests of our formalism using the $\Delta\Sigma(R)$ signal from a truncated NFW profile (see that section for details). The blue solid lines in \fig{fig:nfw} shows $\Sigma(R)$ (top-panel) and $\Delta\Sigma(R)$ for a truncated NFW profile, as well as these same quantities for a point-mass with the same total mass as the truncated NFW profile (orange-dashed lines). For the point-mass case, $\Sigma(R)$ is simply a delta function at $R=0$, while $\Delta\Sigma(R)\propto 1/R^2$. This plot demonstrates the point that 1-halo  contributions with very different scale dependence in $\xigm(r)$ and therefore $\Sigma(R)$ have very similar scale dependence in $\Delta\Sigma(R)$ on all but the smallest scales. This is why marginalizing over a point-mass contribution can effectively account for an uncertain one-halo contribution.

Of course, by marginalizing over $B$ we lose some information, which will result in a loss in constraining power. However, 
we believe this is well justified, since the physical scales informing our model are now 
well-controlled. Assuming that biases in the $\xigm(r)$ prediction increase at smaller physical scales, accounting for the non-local contribution in this way should allow for the robust use of smaller scales in the measurement than if the non-local contribution is ignored. As discussed above, if one does have a motivated prior on the potential size of biases in $\xigm(r)$ at small scales, that information can be naturally included, and the loss in constraining power will be reduced.

\subsection{Analytic marginalization of the enclosed mass contribution}\label{sec:anmarg}

We have described in \sect{sec:theory} how uncertainty in the model prediction for $\Delta\Sigma(R>\rmin)$ that arises from uncertainty in the model prediction for \xigm(r<\rmin) can be accounted for my marginalizing over a term with $1/R^2$ dependence (\eqn{eq:pointmass}). The simple form of this contamination model (e.g.\ the scale dependence is not dependent on cosmology or the lens galaxy properties) makes this term suitable for an analytic marginalization approach (see e.g.\ \citealt{bridle02}). The likelihood desired for our parameter estimation is $P(\Delta\Sigma^{\mathrm{obs}}(R)|\Delta\Sigma^{\mathrm{model}}(R))$
where, as in \eqn{eq:pointmass}, $\dsigr^{\mathrm{model}}$ is the prediction based on a model for $\xigm(r)$ that is correct only for scales $r>\rmin$. This likelihood must be marginalized over the unknown constant $B$, via
\begin{equation}
    P(\dsigobsr|\dsigmodr) = \int \dx{B} P(\dsigobsr | \dsigmodr, B).
\end{equation}
In the case that $\dsigobsr$ is Gaussian distributed with covariance matrix \mathbfss{C}, and we have a Gaussian prior on $B$ with mean zero and width $\sigma_B$, one can show that \citep{bridle02}
P(\dsigobsr|\dsigmodr) is also Gaussian distributed with covariance matrix 
\begin{equation}
    \mathbfss{N} = \mathbfss{C} + \sigma_B^2\vec{x}\vec{x}^{\intercal}\label{eq:anmarg1}
\end{equation}
where $\vec{x}$ has elements $x_a = (\rmin/R_a)^2$.

This powerful result means that operationally, in order to marginalize over the free parameter $B$, we need only perform this simple operation on the original covariance matrix $\mathbfss{C}$, rather than explicitly sampling over possible values of $B$ in e.g. an MCMC chain. 

In the case that we want to use an ``uninformative'' or very wide prior on $B$ (i.e. very large $\sigma_B$), $\mathbfss{N}$ may become close to singular, which will be problematic when numerically calculating $\mathbfss{N}^{-1}$ which is required to compute the Gaussian likelihood. We can circumvent this issue by using the Shermann-Morrison formula to directly calculate $\mathbfss{N}^{-1}$
\begin{align}
    \mathbfss{N}^{-1} &= (\mathbfss{C} + \sigma_B^2\vec{x} \vec{x}^{\intercal})^{-1}\\
    &= \mathbfss{C}^{-1} - \frac{\mathbfss{C}^{-1}\vec{x}\vec{x}^{\intercal}\mathbfss{C}^{-1}}{\vec{x}^{\intercal}\mathbfss{C}^{-1}\vec{x} + \sigma_B^{-2}}.
\end{align}
Indeed with this form we can even use an infinitely wide Gaussian prior on $B$ by taking the limit $\lim{\sigma_B \to \infty}$ in which case 
\begin{equation}
    \mathbfss{N}^{-1} = \mathbfss{C}^{-1} - \frac{\mathbfss{C}^{-1}\vec{x}\vec{x}^{\intercal}\mathbfss{C}^{-1}}{\vec{x}^{\intercal}\mathbfss{C}^{-1}\vec{x}} \label{eq:anmarginfpr}
\end{equation}

\subsection{Relation to ADSD $\Upsilon(R)$}\label{sec:upsilon}

\citet{baldauf10} introduced the Annular Differential Surface Density (ADSD) statistic which they label $\Upsilon(R)$, defined
\begin{equation}
\Upsilon(R; \rmin) = \dsigr - \frac{\rmin^2}{R^2}\Delta\Sigma(\rmin). \label{eq:upsilon}
\end{equation}
This statistic removes the contribution from $R<\rmin$ by effectively using the measured signal at $\rmin$ to estimate the non-local contribution. We can see this by substituting \eqn{eq:pointmass} into \eqn{eq:upsilon}, and observing that terms containing $B$ cancel:
\begin{align}
\Upsilon(R; \rmin) &= \Delta\Sigma^{\mathrm{model}}(R) + B/R^2 \\
&- \frac{\rmin^2}{R^2}\left[ \Delta\Sigma^{\mathrm{model}}(\rmin) + B/\rmin^2 \right] \\
&= \Delta\Sigma^{\mathrm{model}}(R)  - \frac{\rmin^2}{R^2} \Delta\Sigma^{\mathrm{model}}(\rmin)
\end{align}

We demonstrate in \sect{sec:biastest} that this approach has very similar performance to marginalizing over the non-local contribution with infinite prior. Which approach is preferred will likely depend on the details of the analysis. As described in \cite{baldauf10}, a nice feature of the $\Upsilon(R)$ statistic is that is is estimator based, and does not require the introduction of a new free parameter.

Explicitly marginalizing over the point-mass contribution as in our approach (while using analytic marginalization to avoid extra computational cost) allows one to more naturally include a prior, which may allow one to retain more information. We also show that our point-mass marginalization approach can naturally be extended to tomographic measurements where shear-ratio information can be retained, as described in \sect{sec:tomo}.

\subsection{Extension to $\gammat$}

Particularly for photometric lens galaxy samples, $\gammat$ can be a more convenient observable to use than \dsigr\ (and in fact the latter is not a direct observable since a cosmological model must be assumed to calculate $R$ from the angular separation). In the flat-sky and Limber approximations, we can relate \gammat\ to \dsigr\ by integrating over lens and source redshift distributions, $n_{\mathrm{l}}(z)$ and $n_{\mathrm{s}}(z)$:
\begin{equation}
\gamma_t(\theta) = \int \dx{\zlens} \int \dx{\zsource} \nzlens \nzsource
\frac{\Delta\Sigma(R=\theta \times D_{\mathrm{A}}(\zlens), \zlens)}{\sigcrit(\zlens,\zsource)}.
\end{equation}
The $B/R^2$ term in \eqn{eq:pointmass} contributes
\begin{align}
\gamma_t^{\mathrm{pm}}(\theta) &= \int \dx{\zlens} \int \dx{\zsource} \nzlens \nzsource 
\frac{B(\zlens, R=\theta \times D_{\mathrm{A}}(\zlens))}{R^2 \sigcrit(\zlens,\zsource)} \\
&= \theta^{-2} \int \dx{\zlens} \int \dx{\zsource} \nzlens \nzsource \frac{B(\zlens, R=\theta \times D_{\mathrm{A}}(\zlens))}{D^2_{\mathrm{A}}(\zlens)\sigcrit(\zlens,\zsource)}.
\end{align}
where $D_{\mathrm{A}}(\zlens)$ is the angular diameter distance to the lens redshift $\zlens$. Note that the bias in the enclosed mass prediction $B(\zlens, R)$ can now in general depend on the lens redshift and the radius $R$ corresponding to the angular separation $\theta$ at redshift $\zlens$.

Hence for $\gammat$, we can similarly remove the impact of modelling inaccuracies in $\xigm(r)$
at scale $r<\rmin$ by marginalizing over a term with scale dependence $1/\theta^2$, for angular scales $\theta > \rmin/D_{\mathrm{A}}(z_{\mathrm{l,min}})$ where $D_{\mathrm{A}}(z_{\mathrm{l,min}})$ is the distance to the lowest redshift lenses considered (i.e.\ for angular scales corresponding to physical scales greater than $\rmin$ in the lens plane).

\subsubsection{The tomographic case}\label{sec:tomo}

For photometric surveys it is convenient and, given the limited photometric redshift precision, often close to optimal to perform a \emph{tomographic} analysis where lenses and/or source are split into multiple bins in redshift and correlations between all pairs of lens and source redshift bins are used. In this case, our prediction for the tangential shear for lens bin $i$, and source bin $j$ is
\begin{equation}
\gamma_{t,ij}(\theta) = \gamma_{t,ij}^{\mathrm{model}}(\theta) + C_{ij}/\theta^2
\end{equation}
where
\begin{equation}
C_{ij} = \int \dx{\zlens} \int \dx{\zsource} \nzlensi \nzsourcej \frac{B_i(\zlens, R=\theta*D_{\mathrm{A}}(\zlens))}{D^2_{\mathrm{A}}(\zlens)\sigcrit(\zlens,\zsource)}.
\end{equation}
If the lens redshift distribution is sufficiently narrow, or $B_i(\zlens, R)/D^2_{\mathrm{A}}(\zlens)$ evolves with redshift sufficiently slowly across the width of the lens redshift bin, then we can make the approximation 
\begin{align}
C_{ij} &\approx B_i \int \dx{\zlens} \dx{\zsource} \nzlensi \nzsourcej  \sigcrit^{-1}(z_{\mathrm{l},j},\zsource) D^{-2}_{\mathrm{A}}(\zlens) \\
&\equiv B_i \beta_{ij} \label{eq:narrowlens2}
\end{align}
i.e.\ only a single free parameter $B_i$ is required for each lens redshift bin (rather than a free parameter $C_{ij}$ for each lens-source redshift bin pair), with the impact on each lens-source pair modulated by the effective inverse \sigcrit, $\beta_{ij}$. We'll call this the \emph{narrow lens bin assumption}. Of note here is that if we can make this narrow lens bin assumption, then we can extract shear-ratio information from the enclosed mass term, without any assumption about the amplitude of that mass. In effect, we gain constraining power on the relative sizes of the $\beta_{ij}$, which contain geometric information through their dependence on the angular diameter distances which enter \sigcrit\ (see \eqn{eq:sigcrit}), and thus information on cosmological parameters (e.g. \citealt{jain03,taylor07,miyatake17}) and nuisance parameters quantifying e.g. photometric redshift uncertainties (e.g. \citealt{heymans12,prat18}). We note that one can of course choose the width of the lens redshift bins 
in order to attempt to satisfy this narrow lens bin assumption. The success of this approach will depend on whether the lens galaxy redshift uncertainties allow the construction of sufficiently narrow redshift bins. 

In the tests below (\sect{sec:simpletests}, \sect{sec:cosmotests}), we do not explore realistic cases of the redshift evolution of $B_i(\zlens, R)/D^2_{\mathrm{A}}(\zlens)$, which would require realistic galaxy simulations, and is beyond the scope of this work. We instead focus on demonstrating the usefulness of this formalism in idealized cases where the narrow lens bin assumption can be safely assumed.

\subsubsection{Analytic marginalisation}\label{sec:gtanmarg}

In the case that we have $\nlens$ lens redshift bins and $\nsrc$ source redshift bins we need a model for the full length-$N_d$ tangential shear vector (i.e. the concatenation of all angular scales for all lens and source redshift bin pairs) $\vec{\gamma}_t = 
\left[\vec{\gamma}_{t,00},..., \vec{\gamma}_{t,0\nsrc},...,\vec{\gamma}_{t,\nlens\nsrc}\right]$. 

In the case that we can make the narrow lens bin assumption in \eqn{eq:narrowlens2}, we can use the following form:
\begin{equation}
    \gamma_{t}(\theta) = \gamma_{t}^{\mathrm{model}}(\theta) + B_i \beta_{ij} \theta^{-2}\label{eq:gtpm1}
\end{equation}
where $i$ and $j$ are the lens and source redshift, and again $\vec{\gamma_t}^{\mathrm{model}}$ is based on a $\xigm$ prediction that is accurate only down to some scale $\rmin$. It is useful to write this in vector notation:
\begin{equation}\label{eq:pmsig}
    \vec{\gamma_{t}} = \vec{\gamma_{t}}^{\mathrm{model}} + \sum_i^{\nlens} B_i \vec{t}_i
\end{equation}
where
\begin{equation}\label{eq:ti}
    (\vec{t}_i)_a = \begin{cases}
    0 & \text{if lens redshift bin for element $a$ is not $i$}\\
    \beta_{ij}\theta_a^{-2} & \text{otherwise}
    \end{cases}
\end{equation}
where $j$ is the source redshift bin and $\theta_a$ is the angular separation for element $a$ of the full datavector. We note that in the above, and throughout, we use $i$ and $j$ as redshift bin labels rather than vector indices; $\vec{t}_i$ in \eqn{eq:gtpm1} does not represent element $i$ of a vector $\vec{t}$, rather one of a set of $\nlens$ vectors. When we do provide a piece-wise definition of a vector we use an index $a$ as in \eqn{eq:ti}.

Analytic marginalization over all $B_i$ can again be performed by updating the covariance matrix to $\mathbfss{N}$ given by
\begin{equation}
    \mathbfss{N} = \mathbfss{C} + \sum_i \sigma_{B_i}^2 \vec{t}_i\vec{t}_i^{\intercal}
    \label{eq:matBi}
\end{equation}
where $\mathbfss{C}$ is the original $\gamma_t$ covariance and $\sigma_{B_i}$ is width of the Gaussian prior on $B_i$. 

We can also write $\mathbfss{N}$ in the form
\begin{equation}
    \mathbfss{N} = \mathbfss{C} + \mathbfss{U}\mathbfss{U}^{\intercal}
\end{equation}\label{eq:UV}
where $\mathbfss{U}$ is a $N_d \times \nlens$ matrix with $i$th column $\sigma_{B_i}\vec{t}_i$, and $N_d$ is the number of elements in the datavector. We will refer to $\mathbfss{U}$ as a \emph{template matrix} since its columns are template modes to be marginalized over. We can use the Woodbury matrix identity (the generalization of the Sherman-Morrison formula introduced in \sect{sec:anmarg}) to get the inverse:
\begin{equation}
    \matn^{-1} = \matc^{-1} - \matc^{-1}\mathbfss{U}(\mathbfss{I} + \mathbfss{U}^{\intercal}\matc^{-1}\mathbfss{U})^{-1}\mathbfss{U}^{\intercal}\matc^{-1}.
    \label{eq:woodbury}
\end{equation}
where $\mathbfss{I}$ is the identity matrix. 

Again, we may want to consider the case where we allow maximal freedom in the model by taking the limit $\sigma_{B_i} \to \infty$. In this case \eqn{eq:woodbury} reduces to 
\begin{equation}
    \matn^{-1} = \matc^{-1} -
    \matc^{-1}\mathbfss{V}(\mathbfss{V}^{\intercal}\matc^{-1}\mathbfss{V})^{-1}\mathbfss{V}^{\intercal}\matc^{-1}.
\end{equation}
where \mathbfss{V} is a $N_d \times \nlens$ matrix with $i$th column $\vec{t}_i$.

If we cannot make the narrow lens bin assumption, then we have 
\begin{equation}
    \vec{\gamma_{t}} = \vec{\gamma_{t}}^{\text{model}} + \sum_{i=1}^{\nlens}\sum_{j=1}^{\nsrc}C_{ij} \vec{t}_{ij}
    \label{eq:pmnosig}
\end{equation}
where we now use $ij$ to label the lens-source redshift bin \emph{pair}, and
\begin{equation}
    (\vec{t}_{ij})_a = \begin{cases}
    \theta_a^{-2} & \begin{aligned} &\text{if the lens-source redshift bin pair}\\
    &\text{for element $a$ is $ij$}
    \end{aligned}\\
    0 & \text{otherwise}
    \end{cases}
\end{equation}
We can again marginalize over the free parameters $C_{ij}$ analytically, transforming the covariance matrix $\mathbfss{C}$ according to \eqn{eq:UV}.
In this case $\mathbfss{U}$ is a $N_d \times N_p$ matrix, where $N_p=\nlens \times \nsrc$, the total number of lens-source redshift bin pairs. The $p$th column (where $p=i\times\nlens + j$) is given by $\sigma_{C_{ij}}\vec{t}_{ij}$ where $\sigma_{C_{ij}}$ is the width of the Gaussian prior on $C_{ij}$.

\subsection{Simple Tests}\label{sec:simpletests}

In this section we test the the above formalism by using it in parameter estimation on galaxy-galaxy lensing datavectors with reasonable one-halo contamination. For all tests where we use analytic marginalization we use the infinite prior case. For the point-mass marginalization, this means we make no assumption about $\xigm(r)$ below $\rmin$. 

We first describe our simulated datavector, which is used here and in \sect{sec:cosmotests}. We generate a galaxy-galaxy lensing and galaxy clustering datavector based on that used in the DES Year 1 analyses of \citet{methodpaper,keypaper}. 
The galaxy clustering part of the datavector is not used in this section, but is used in \sect{sec:cosmotests}. The lens sample has 5 redshift bins spanning a range of $0.15-0.9$ in redshift. The source sample has 4 redshift bins, roughly spanning a range $0.2-1.3$ in redshift (see Figure 1 of \citet{keypaper} for more details).
Both galaxy-galaxy lensing and galaxy clustering signals are generated according to a linear bias model, with values of the galaxy bias, $b_i = \left[1.45, 1.55, 1.65, 1.8, 2.0\right]$. Simulated measurements are generated for 20 log-spaced bins between angular scales $2.5$ and \SI{250}{\arcminutes}. Again following \citet{keypaper,methodpaper}, in our linear bias model the galaxy-galaxy and galaxy-matter power spectra for redshift bin $i$ are in fact generated as $P_{\mathrm{gg}}=(b_i)^2 P_{\mathrm{nl}}$ and $P_{\mathrm{gm}}=b_i P_{\mathrm{nl}}$, where $P_{\mathrm{nl}}$ is the nonlinear matter power spectrum calculated using \textsc{halofit} \citep{smith03,takahashi2012}. Throughout, we use the \textsc{CosmoSIS}\footnote{\url{https://bitbucket.org/joezuntz/cosmosis/}}  package \citep{cosmosis} for theory predictions and parameter inference, implementing in custom modules the 1-halo contamination and Gaussian covariance calculation described below, as well as our analytic marginalization scheme.

We generate a Gaussian covariance matrix corresponding to this datavector which has roughly DES Year 5 
statistical power - we assume an area of $\SI{5000}{\deg^2}$, a lens galaxy number density of $\left[ 0.013,  0.034,  0.051,  0.030,  0.0088 \right]$ galaxies per square arcminute, a source galaxy number density of 2 per square arcminute for each redshift bin (i.e. totalling 8 source galaxies per square arcminute), and $\sigma_e=0.4$ (the total ellipticity dispersion) for all source redshift bins. 

We add to the galaxy-galaxy lensing simulated datavector a simple 1-halo contribution (on top of the linear bias term already present). For each lens redshift bin, based on the fiducial linear bias values above, and the mean redshift of the bin, we calculate a fiducial halo mass $M_{200}$ value following \citet{tinker10} and concentration, $c$ following \citet{duffy08}. The 1-halo contribution is then calculated as the tangential shear from a truncated NFW profile with mass $M_{200}$ and concentration $c$ at the mean redshift of the lens redshift bin only - meaning we can safely make the narrow lens assumption for this contamination term. For convenience, we use the smoothly truncated NFW density profile in Equation A.3 of \citet{baltz09}, with $\tau=2$, since this allows for analytic calculation of $\Delta\Sigma(R)$.

We note that our model for the galaxy-matter correlation function, which is the sum of a linear bias 2-halo contribution, and a one-halo term, is not very realistic. We do not include satellite galaxies in the model, or a realistic distribution of halo mass and concentration. It is unlikely to be accurate in the \emph{transition} regime, where the one and two-halo contributions are of comparable size. Additionally, our 2-halo term does not have a cut-off at small scales to account for halo exclusion (e.g. \citealt{smith07}) so may make too large a contribution here. We stress that this model, while of limited realism, is well-suited to demonstrate the usefulness of the techniques presented here, where we marginalize over the impact of the 1-halo contribution without making assumptions about its functional form or even amplitude. We defer the study of more realistic one-halo contamination to future work using galaxy simulations, focusing here instead on proof-of-concept type tests. 

Unless otherwise stated, we throughout impose minimum angular scales in our simulated datavector corresponding to $\SI{4}{\hinv\megaparsec}$ at the mean redshift of the lens redshift bin, which equate to $\left[21.5,13.5,10.0,8.0,7.0 \right]\ \mathrm{arcminutes}$ for the five lens redshift bins.

\subsubsection{Recovering the linear galaxy bias from $\gammat$}\label{sec:biastest}

\begin{figure}
\includegraphics[width=\columnwidth]{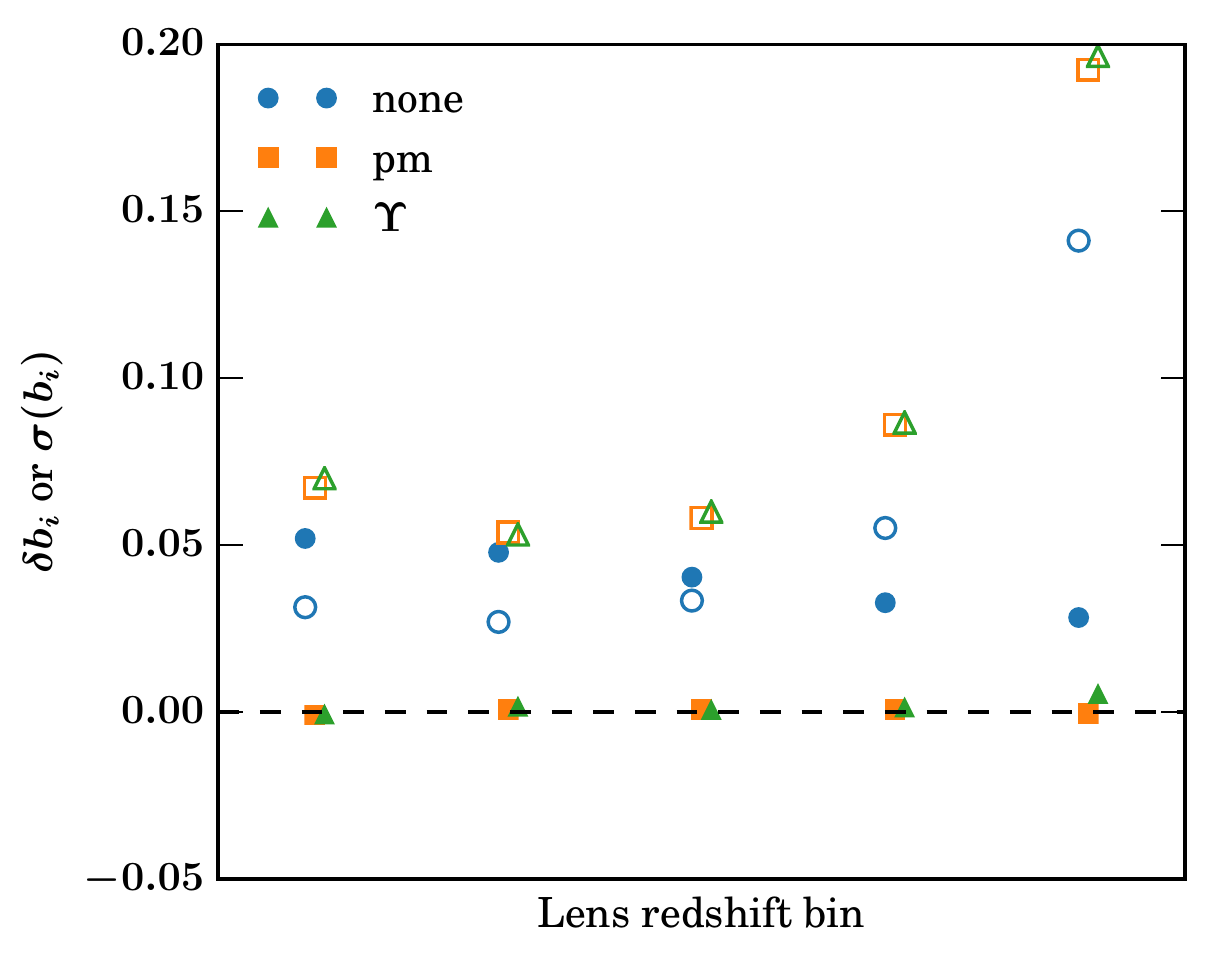}
\caption[]{The bias (filled markers) and $1-\sigma$ unertainty (unfilled markers) on the inferred linear galaxy bias values from a DES-like galaxy-galaxy lensing datavector with five lens redshift bins (described in \sect{sec:simpletests}) at fixed cosmology. The blue markers show the case where no marginalization is performed to account for the 1-halo contribution and hence the recovered values are somewhat biased. The orange markers show the case where the analytic marginalization scheme in \sect{sec:gtanmarg} is used. Green markers use the ADSD statistic and recover essentially the same constraints as the point-mass marginalization approach. }
\label{fig:biastest}
\end{figure}

As a first simple step we test the recovery of the linear bias, $b_i$ for each lens redshift bin $i$, from the galaxy-galaxy lensing signal only.  We fix all other parameters fixed and analyse the datavector with and without using our analytic marginalization scheme to account for the 1-halo contamination. The solid markers in  \fig{fig:biastest} shows the bias in the recovered $b_i$, calculated as $\left< b_i \right> - b_{i,\text{true}}$ where $\left< b_i \right>$ is the mean of the marginalized posterior and $b_{i,\text{true}}$ is the galaxy bias value used to generate the datavector. For the blue circles marginalization over the point-mass contribtution was not performed, and hence biased values for the galaxy bias are recovered, while for the orange markers point-mass marginlaization was included, and the correct galaxy bias values are recovered. The corresponding open markers indicate the $1\sigma$ uncertainty on the recovered linear bias parameter, demonstrating that as expected, there is some degradation in constraining power when using the point-mass marginalization scheme.

We also implement the analogous ADSD statistic for $\gammat$, 
\begin{equation}
    \Upsilon(\theta)=\gamma_t(\theta) - \left(\frac{\theta}{\thetamin}\right)^2\gamma_t(\thetamin).
\end{equation}
This shows very similar performance to the point-mass marginalization approach.
For simplicity, we use as $\gamma_t(\thetamin)$ simply the smallest angular bin in the measurement remaining after applying the scale cuts. When the ADSD statistic has been used on data, an estimate of $\Delta\Sigma(\rmin)$ has typically been made by fitting a power-law over a range of scales around $\rmin$ (e.g. \citealt{mandelbaum13}). We do not attempt to compare to this more complex approach here, although we note that analogous information could be added to the point-mass marginalization.

\subsubsection{Shear-ratio information}\label{sec:dztest}

\begin{figure*}
\includegraphics[width=14cm]{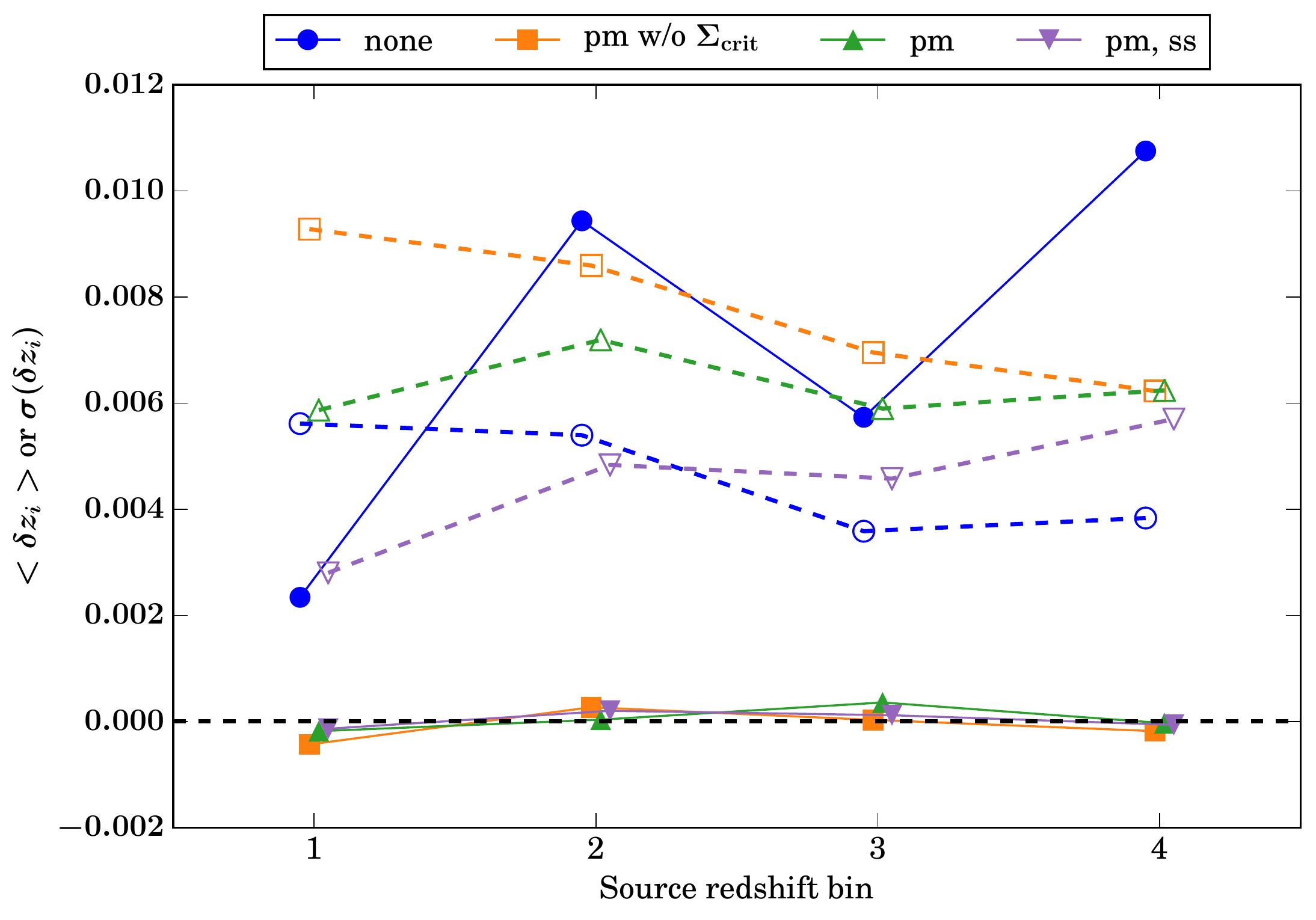}
\caption[]{
The bias (filled markers, solid lines) and uncertainty (unfilled markers, dashed lines) on the inferred value of $\delta z_i$, the shift in the redshift distribution for source redshift bin $j$, for different small scale $\gammat$ treatments. The datavector is based on a DES Year 1 galaxy-galaxy lensing datavector with DES Year 5-like uncertainties. `none' indicates no point-mass marginalization is performed and small-scales are removed. `pm w/o $\sigcrit$' indicates point-mass marginalization is performed for each lens-source redshift bin pair independently (i.e. no shear-ratio relation is assumed). `pm' indicates point-mass marginalization is performed for each lens redshift bin independently, retaining shear-ratio information.
`pm, ss' indicates both point-mass marginalization and small-scale marginalization are performed, the approach that retains the most information while still being unbiased.}
\label{fig:dztest}
\end{figure*}

As described in \sect{sec:tomo}, if we make the narrow lens bin assumption we can straightforwardly retain the shear-ratio information in the point-mass term. One way to demonstrate this is to allow some freedom in the redshift distributions of the source redshift distributions, and test our constraining power on these distributions. Shear-ratio measurements have already been used for this application by \citet{pratsanchez18} who extracted competitive constraints on these shift parameters using DES Year 1 galaxy-galaxy lensing measurements. 
Using the same contaminated galaxy-galaxy lensing datavector described above, we allow a simple shift $\delta z_j$ for source redshift bin $j$. For this test we keep the linear bias and cosmological parameters fixed to their true values. We produce constraints on the $\delta z_j$ with three different modeling approaches, introduced here with the same labelling used in \fig{fig:dztest}:
\begin{enumerate}
    \item{`none': we do not marginalize over the point-mass contribution and hence expect this approach to produce the tightest, but also biased constraints.}
    \item{`pm w/o \sigcrit': we analytically marginalize over the point-mass contribution but allow a fully independent contribution for each lens-source redshift bin pair (following \eqn{eq:pmnosig}, thereby not retaining shear-ratio information from the point-mass contribution.}
    \item{`pm': we analytically marginalise over an independent point-mass contribution for each lens redshift bin i.e. assume a perfectly correlated contribution to all source redshift bins for a given lens bin (following \eqn{eq:pmsig}).}
\end{enumerate}

\fig{fig:dztest} shows the mean of the posterior on $\delta z_j$ (solid markers, solid lines) and its $1-\sigma$ uncertainty (open markers, dashed lines) for the three cases above (blue circles, orange squares, and green triangles respectively). We see again that there is a degradation in constraining power when performing the point-mass marginalization with either method (ii) or (iii), but that this degradation is smaller for the case (iii) where shear-ratio information in the point-mass contribution is retained, particularly for the two lower source redshift bins.


\section{Making use of all measured scales in the tangential shear}\label{sec:smallscalemarg}

We can extend the above formalism to make use of all measured scales in our $\gamma_t(\theta)$ measurement. As above, we assume that for each lens redshift bin, $i$, there is an angular scale $\thetamin^i$ which corresponds to a physical scale in the lens plane $\rmin$, below which we do not have a trustworthy model for the galaxy-matter correlation function $\xigm (r)$. We therefore cannot make a reliable model prediction for $\gamma_t(\theta<\thetamin$) (even if a point-mass contribution were marginalized over). However if we can make the narrow lens redshift bin assumption, we do know how the relative amplitudes of the different lens-source bin combinations for a given lens redshift bin are related i.e.
\begin{equation}
    \gamma_{t,ij}(\theta<\thetamin)/\gamma_{t,ij'}(\theta<\thetamin) = \beta_{ij}/\beta_{ij'}.
\end{equation}\label{eq:shearratio}

For any measured scale we can therefore write down our model for lens redshift bin $i$ and source redshift bin $j$ as:
\begin{equation}
    \gamma_{t,ij}(\theta) = \gamma_{t,ij}^{\mathrm{model}}(\theta) + \beta_{ij}B_i/\theta^2 +
    \beta_{ij}D_i(\theta).
    \label{eq:gt_all_scales}
\end{equation}
The second term is the point-mass contribution described in \sect{sec:pm} and is not included for scales $\theta<\thetamin^i$.
We have now added a third term containing a function $D_i(\theta)$ that is zero for $\theta>\thetamin^i$, and allowed to vary freely for scales $\theta<\thetamin^i$. 
Physically, for $\theta<\thetamin^i$ we are allowing the value of $\Delta\Sigma(R=\theta D_l)$ to vary freely for $\theta<\thetamin^i$, while enforcing that it takes the same value for a given angular scale and lens redshift bin, and hence is simply modulated in $\gamma_{t,ij}$ by $\beta_{ij}$. We refer to marginalization over the $D_i(\theta)$ as \emph{small-scale marginalization}. 

One may ask if we are introducing too much freedom in the model, since the point-mass contribution at large scales is  determined by the density profile at small scales, which we are also now marginalizing over. However, the density profile is never fully determined down to zero, since the shape noise on the measurement diverges in the limit of zero angular separation. Using some parameterization for the projected density profile down to zero would allow the information from these small scales to constrain the point-mass contribution to larger scales, potentially reducing degradation in the constraints when marginalizing over the point-mass term. We leave further investigation of this approach for future work.


Analytic marginalization is again extremely useful here, since we may want to marginalize over 10s or 100s of $D_i(\theta)$ values. In order to perform analytic marginalization, it is again useful to recast in vector notation, with the full $\gammat$ datavector given by 
\begin{equation}
    \vec{\gamma_{t}} = \vec{\gamma_{t}}^{\mathrm{model}} + \sum_i^{\nlens} B_i \vec{t}_i + \sum_i^{\nlens} \sum_k^{N_{\theta}^i}
    D_{ik} \vec{s}_{ik}
    \label{eq:gtsmallscale}
\end{equation}

where $N_{\theta}^i$ is the number of angular bins for lens bin $i$ with $\theta<\thetamin^i$, and 
\begin{equation}
    (\vec{s}_{ik})_a = \begin{cases}
    \beta_{ij}\Theta(\theta_a, \thetamin^i) & 
    \begin{aligned} 
    & \text{if lens redshift bin and angular bin}\\ 
    & \text{for element $a$ are $i$ and $k$ respectively}
    \end{aligned}\\
    0 & \text{otherwise}
    \end{cases}
\end{equation}
where $j$ is the source redshift bin for datavector element $a$ and 
\begin{equation}
    \Theta(X, Y) = \begin{cases}
    1 & \text{if $X<Y$}\\
    0 & \text{otherwise.}
    \end{cases}
\end{equation} 

We also update the definition of $\vec{t}_i$ such that the point-mass contribution is not marginalized over for scales $\theta<\thetamin^i$:
\begin{equation}
    (\vec{t}_i)_a = \begin{cases}
    0 & \text{if lens redshift bin for element $a$ is not $i$}\\
    \beta_{ij}\theta_a^{-2}\Theta(\thetamin,\theta_a) & \text{otherwise}
    \end{cases}
\end{equation}

The $D_{ik}$ in \eqn{eq:gtsmallscale} are the set of free parameters we introduce to marginalize over the density profile at small scales. We note again that $i$, $j$ and $k$ in the above are not vector or tensor indices, but rather labels for lens redshift bin and angular bin respectively i.e. $D_{ik}$ and $s_{ik}$ are from sets of $\sum_i^{\nlens} N_{\theta}^i$ scalars and vectors respectively.
As in the case of the point-mass contribution, we can perform this marginalization analytically by updating the $\gammat$ covariance matrix according to \eqn{eq:UV}, where now $\mathbfss{U}=(\mathbfss{U}_{pm}|\mathbfss{U}_{ss})$, the concatenation of the template matrices for the  point-mass marginalization and the small-scale marginalization over the $D_i(\theta)$ described in this section. $\mathbfss{U}_{ss}$ is a matrix with dimensions $N_d \times N_q$, where $N_q=\sum_i^{\nlens} N_{\theta}^i$, the total number of datapoints with $\theta<\thetamin^i$. It has columns $\sigma_{D_{ik}}\vec{s}_{ik}$ where $\sigma_{D_{ik}}$ is the Gaussian prior width for the free parameter $D_{ik}$.

We repeat the test in \sect{sec:dztest} in which we analyse a $\gammat$ datavector, which contains unmodelled contamination by a 1-halo term, while allowing shifts in the source bin redshift distributions $\delta z_j$. To increase the usefulness of the small-scale marginalization described here, we extend the simulated measurements down to smaller scales - adding another 10 logspaced angular bins between 0.25 and 2.5 arcminutes. When not using the small-scale marginalization scheme, these extra scales are removed by the scale cuts, so have no bearing. 

The results are again plotted in \fig{fig:dztest}. The purple downwards facing triangles represent the case where both point-mass and small-scale marginalization are used. Compared to the ``pm" case where only point-mass marginalization is used, the inferred $\delta z_i$ values remain unbiased, but uncertainties are significantly  reduced. In the two lowest redshift bins, the extra information reduces statistical uncertainties to below those in the ``none" (no point-mass or small-scale marginalization) case.
We conclude that this double-pronged approach of marginalization over both the point-mass contribution, and the underlying signal at too-small-to-model scales, is the most successful at recovering unbiased shear-ratio information.

\section{Tests of cosmological parameter estimation}\label{sec:cosmotests}


\begin{figure*}
\includegraphics[width=\columnwidth]{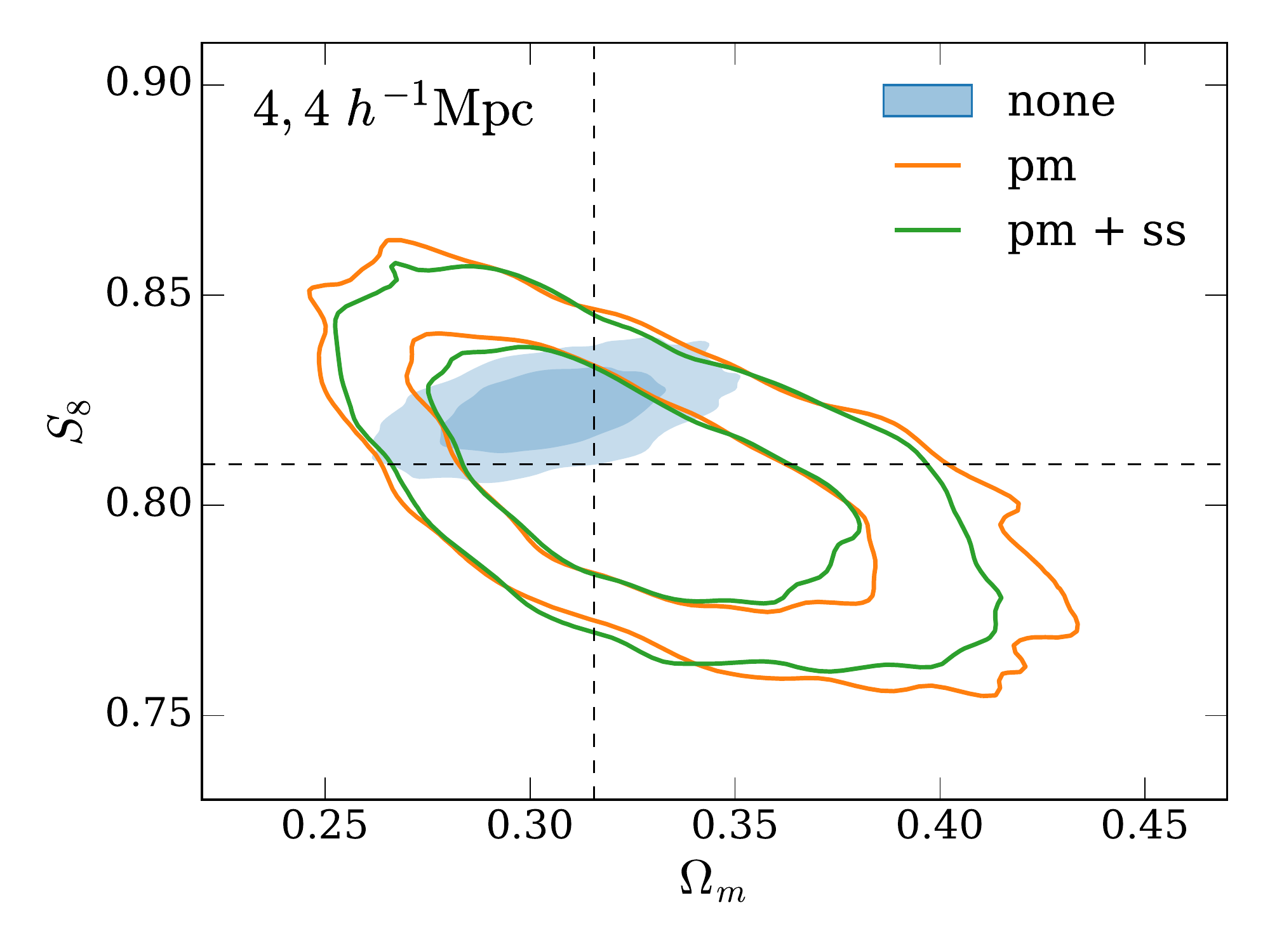}
\includegraphics[width=\columnwidth]{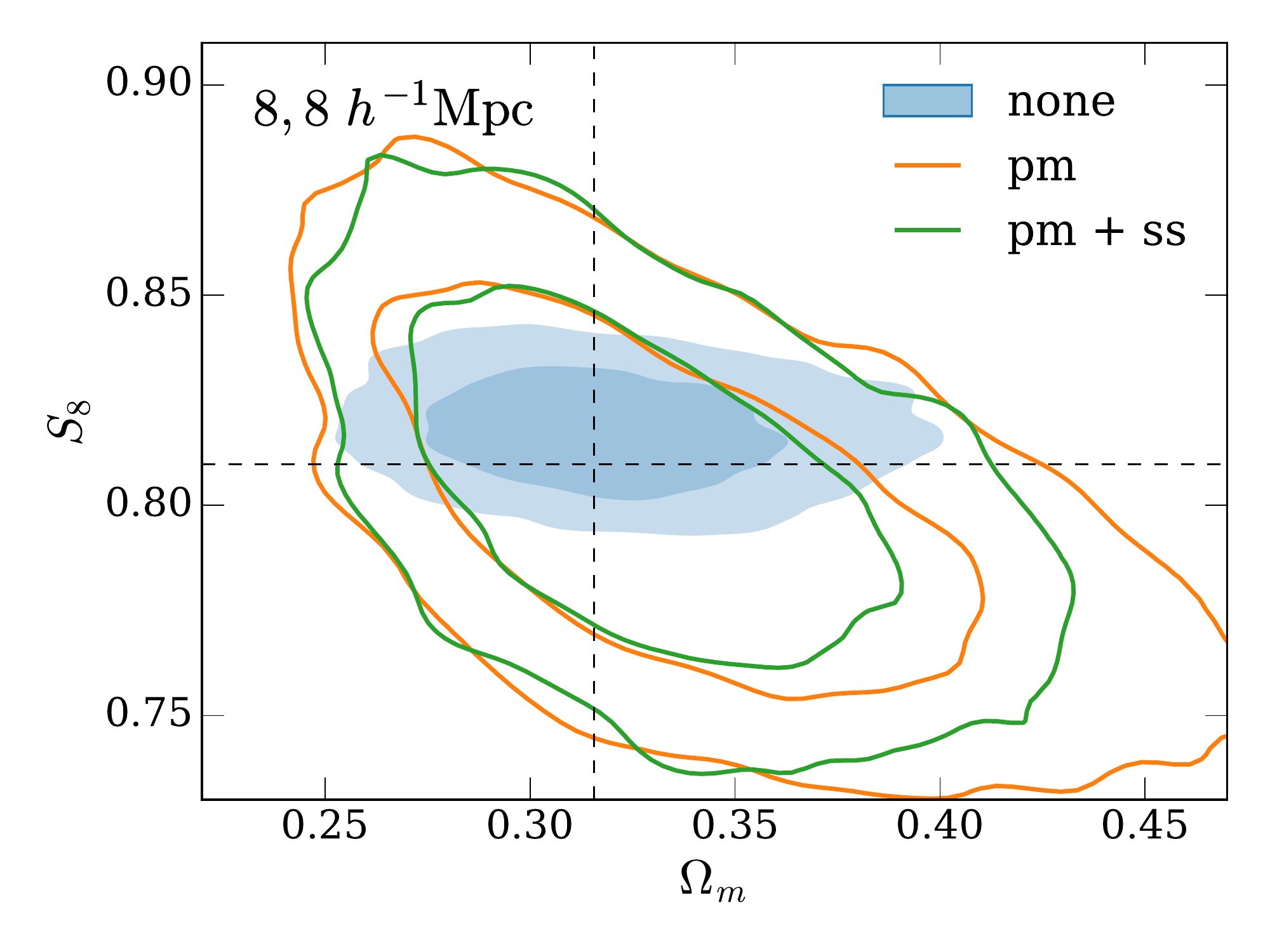}
\includegraphics[width=\columnwidth]{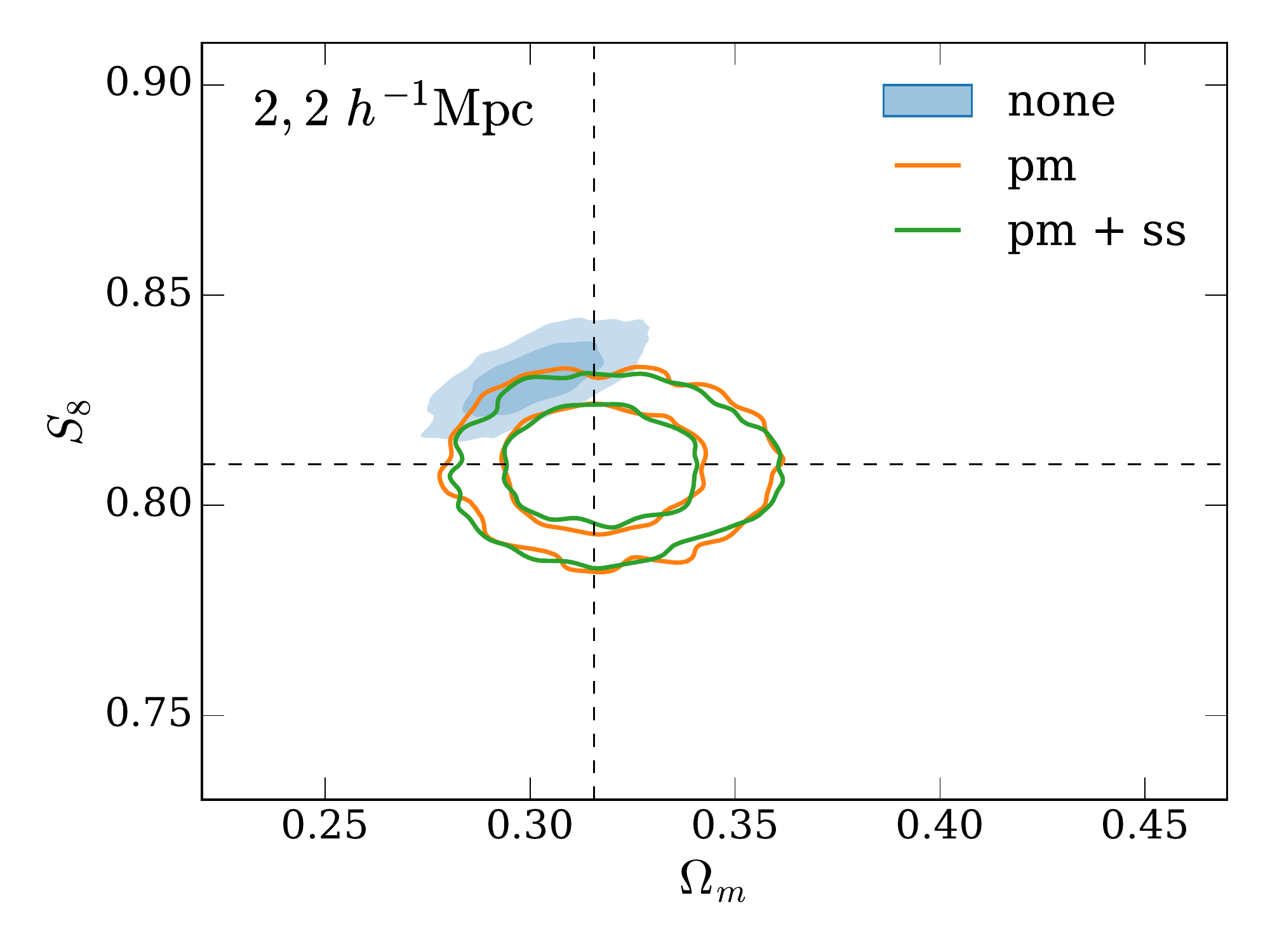}
\includegraphics[width=\columnwidth]{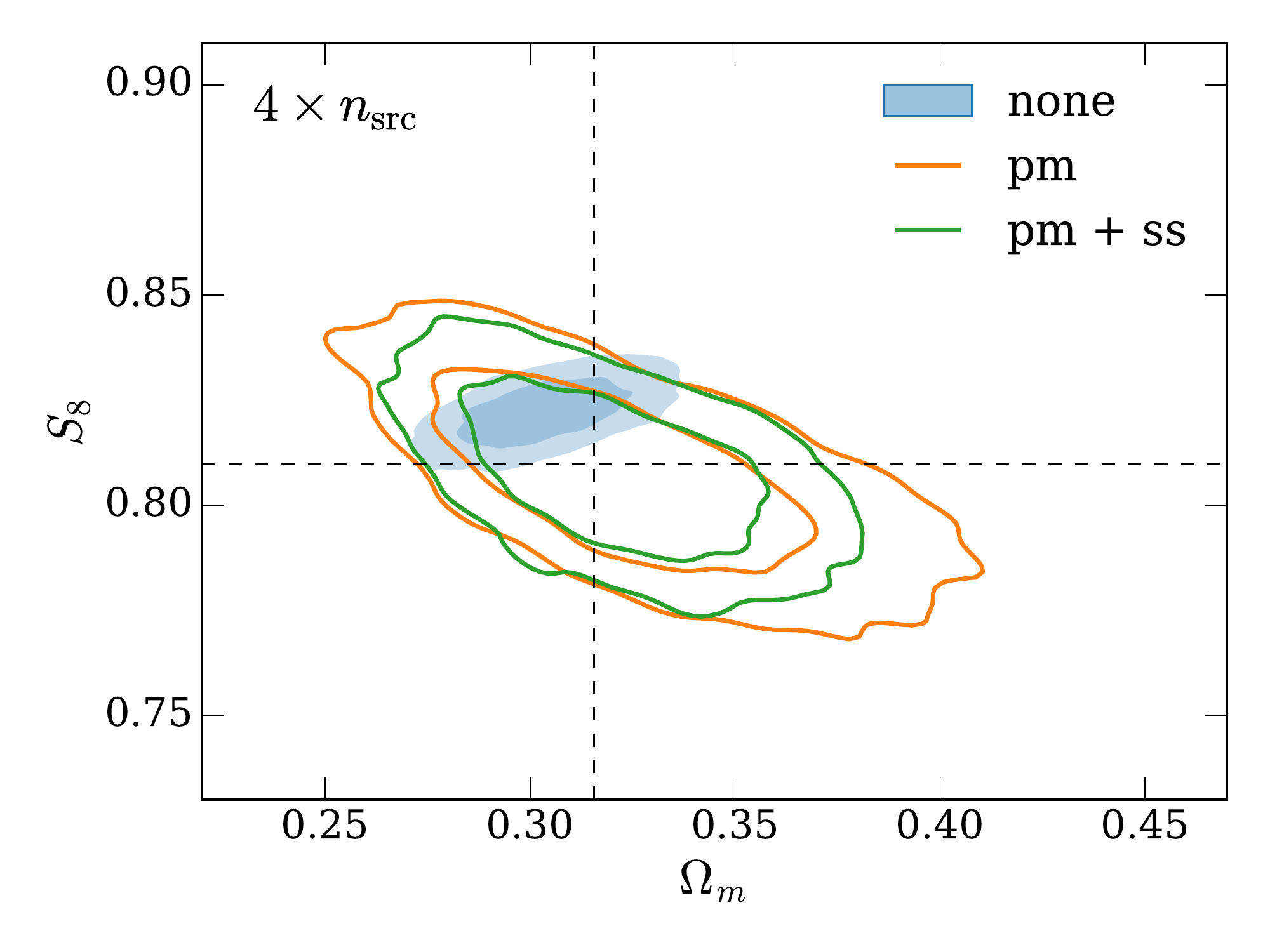}
\caption[]{All panels show constraints on $\om$ and $S_8=\sigma_8(\om/0.3)^{0.5}$ for a DES-like galaxy-galaxy lensing and galaxy clustering analysis. The simulated datavector has contamination by an un-modeled one-halo term in the galaxy-galaxy lensing signal (described in \sect{sec:simpletests}). In addition to $\om$ and $\sigma_8$, the Hubble constant $H_0$, and a linear bias for each redshift bin are varied (see \sect{sec:cosmotests} for details).  The true values (i.e. those used to generate the datavector) are shown as the grey dashed lines. The three sets of contours represent the three modeling approaches described in \sect{sec:cosmotests}. Blue solid contours result from using neither point-mass or small-scale marginalization. The orange outlined contours use point-mass but not small-scale marginalization. The green outlined contours use both point-mass and small-scale marginalization. 
The top-left panel is for our fiducial setup described in \sect{sec:simpletests}. In the top-right (bottom-left) panels we use larger (smaller) minimum scale cuts corresponding to $\SI{8}{\h^{-1}\megaparsec}$ ($\SI{2}{\h^{-1}\megaparsec}$) in the lens plane for both galaxy-galaxy lensing and clustering. In the bottom-right panel, a source galaxy number density 4 times higher than the fiducial setup is assumed.
}
\label{fig:omsig}
\end{figure*}

We perform some simple simulated-likelihood tests to show how our analytic marginalization scheme helps with unbiased cosmological inference. We simulate parameter estimation on the joint galaxy-galaxy lensing and galaxy clustering datavector described in \sect{sec:simpletests}.  For each lens redshift bin $i$,  we again cut out angular scales less than $\thetamin^i$  corresponding to $<\SI{4}{\hinv\megaparsec}$ in the lens plane for both the galaxy-galaxy lensing and clustering measurements, except in the case that we retain these scales for the small-scale marginalization scheme described in \sect{sec:smallscalemarg}. In this case angular scales less than $\thetamin^i$ are retained in galaxy-galaxy lensing only, and are only used for the small-scale marginalization scheme, rather than being modeled explicitly.

We analyse the datavector (again using the DES Year 5-like covariance) with three modelling approaches:
\begin{enumerate}
    \item{No marginalization (point-mass or small-scale) is performed}
    \item{only point-mass marginalization is performed}
    \item{both point-mass and small-scale marginalization are performed}
\end{enumerate}

Firstly, we vary only a linear bias parameter for each lens redshift bin, the matter density, \om\ in the range $[0.1,0.9]$, the amplitude of the primordial power spectrum, $A_s$ in the range  $[\SI{0.5e-9}{},\SI{5e-9}{}]$ and $h$ in the range $[0.4,0.95]$ (with the Hubble constant given by  $H_0=\SI{100}{\h\ (\km/\second)/\megaparsec}$). We assume a flat \lcdm\ cosmology with all other cosmological parameters fixed. $\sig$ is recorded in our MCMC chains as a derived parameter.
For our fiducial setup, the resulting constraints on \om\ and $S_8=\sig(\om/0.3)^{0.5}$ are shown in the top-left panel of \fig{fig:omsig} (here and throughout, contours represent the 68\% and 95\% credible intervals). As expected, modelling approach (i) results in the tightest, but biased constraints, since potential contamination by the 1-halo term is not marginalized over. 
Cases (ii) and (iii) recover the true cosmology (indicated by the dashed lines) correctly. 
For this parameter space there is a modest gain in constraining power when using the small-scale marginalizaiton (i.e. the gain in case (iii) over case (ii)), with a $16\%$ decrease in the uncertainty on \om.

In the other three panels of \fig{fig:omsig}, we study how the different modeling approaches perform under variations to our fiducial setup. The top-right and bottom-left panels use larger and smaller minimum scales in both galaxy-galaxy lensing and clustering (corresponding to $\SI{8}{\h^{-1}\megaparsec}$ and $\SI{2}{\h^{-1}\megaparsec}$ in the lens plane), with a couple of trends becoming apparent. Firstly, when using smaller scales, the bias when not performing point-mass marginalization (the contour labelled `none') is increased in significance. Secondly, when using smaller scales, the gain from using the small-scale marginalization relative to point-mass marginalization only is reduced (this is expected since there is simply less information for the small-scale marginalization to reclaim). 

Finally, in the bottom-right panel, we re-compute the covariance matrix with four times the density of source galaxies (keeping our fiducial scale cuts). This is potentially useful to gain an intuition on how the results here may apply to other lensing surveys which have a higher source galaxy density than DES, like the Hyper Suprime-Cam Subaru Strategic Survey\footnote{https://hsc.mtk.nao.ac.jp/ssp/} (HSC, \citealt{mandelbaum18b}), LSST or WFIRST. Increasing the source galaxy number density decreases the shape noise on the galaxy-galaxy lensing measurement, which is the dominant contributer to the covariance on small scales. 
This makes the bias in inferred parameters when no point-mass marginalization is performed (the `none' case) more significant than in the fiducial case. The extra signal-to-noise on small scales also leads to a greater gain in constraining power when using small-scale marginalization compared to point-mass marginalization only, with the $1-\sigma$ uncertainty on $\om$ reduced by 23\%.

An impression one may take from \fig{fig:omsig} is that the decrease in constraining power when using the point-mass marginalization (orange and green outlined contours) is rather large compared to the parameter bias when not using it (blue solid contour). We note that the size of this bias here is a direct result of the highly simplified, order-of-magnitude model we've chosen for inaccuracy in the $\xigm(r)$ model (contamination by a simple 1-halo term), our chosen scale cuts (e.g. using smaller scales would result in greater bias), and survey properties. In reality, for a given galaxy sample and observational setup, biases could be much larger or smaller. Furthermore, unlike in the test presented here, one would be unlikely to use the same minimum scale cut in the two approaches - a more conservative minimum scale would likely be required when not marginalizing over the point-mass to meet some requirement on the ratio between inferred parameter bias and uncertainty. We further note that informative priors on the point-mass contribution can be naturally included in our framework and would reduce the degradation when including the point-mass marginalization.

\begin{figure*}
\includegraphics[width=14cm,trim={2cm 1cm 2cm 1cm},clip]{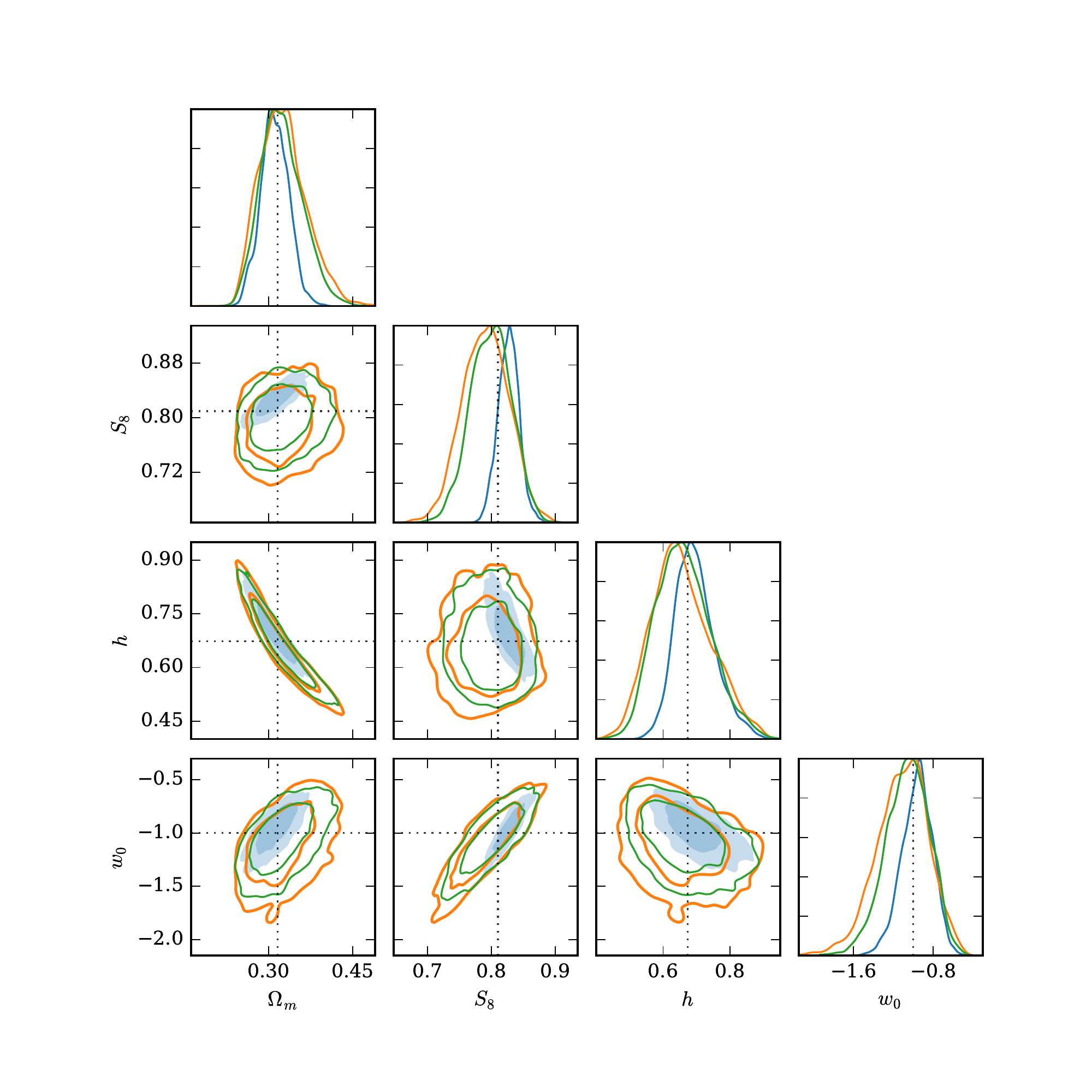}
\caption[]{Simulated constraints on cosmological parameters from a DES Year 5-like galaxy-galaxy lensing and galaxy lustering analysis (see \sect{sec:cosmotests} for details). The simulated datavector has contamination by an un-modeled one-halo term in the galaxy-galaxy lensing signal (described in \sect{sec:simpletests}). All varied cosmological parameters are shown. A linear bias parameter for each lens redshift bin is also marginalized over. The true values (i.e. those used to generate the datavector) are shown as the grey dotted  lines. The three sets of contours represent the three modeling approaches described in \sect{sec:cosmotests}. Blue solid contours result from using neither point-mass or small-scale marginalization. The orange outlined contours use point-mass but not small-scale marginalization. The green outlined contours use both point-mass and small-scale marginalization.}
\label{fig:oms8hw}
\end{figure*}

Next, (returning to our fiducial scale cuts and source galaxy density) we additionally allow $w_0$, the (constant with redshift) dark energy equation of state parameter, to vary from its \lcdm\ value of $-1$, in the range $[-3,-0.33]$. \fig{fig:oms8hw} shows marginalized constraints on \om\, $S_8$, $h$ and $w_0$. Again, modeling approach (i) recovers the tightest constraints, but biases with respect to the truth values are present, with the truth lying outside the $68\%$ credible interval in the $S_8-h$ and $S_8-\om$ planes for example. Again, when using small-scale marginalization, modest gains in constraining power are apparent in most of the 2d projections of the posterior, and the constraint on $w_0$ is improved by $16\%$ with respect to the case when only point-mass marginalization is used.

\section{Discussion}\label{sec:discussion}

We have described and demonstrated a methodology which uses an analytic marginalization approach to target two issues with small scale galaxy-galaxy lensing measurements.
Firstly, the galaxy-galaxy lensing signal measured at physical separation $R$ in the lens plane receives significant contributions from scales $r<R$ in the galaxy-matter correlation function $\xigm(r)$.
We have described how uncertainty in the model prediction for this contribution can be straightforwardly marginalized over by including in the model a $1/R^2$ (for $\dsigr$) or $1/\theta^2$ (for $\gammat$) dependence with free amplitude. We demonstrate that this approach can successfully remove biases in inferred parameters when an un-modeled one-halo contribution is present in the galaxy-galaxy lensing signal, and that this marginalization can be performed analytically, to avoid adding extra sampling parameters to the parameter inference. We note that the approach of \citet{baldauf10} also achieves this goal, although our approach may more naturally allow the use of priors and retention of shear-ratio information in a tomographic analysis.

Secondly, we demonstrate that an analytic marginalization approach can also be used to extract shear-ratio information from small scale galaxy-galaxy lensing measurements that would otherwise be excluded due to modelling uncertainties i.e. those corresponding to physical separation in the lens plane $R<\rmin$
, where $\rmin$ is the smallest physical scale for which a $\xigm(r)$ prediction is accurate.
Again, the use of analytic marginalization allows us include many extra nuisance parameters without having to explicitly sample over them in a Monte Carlo chain, making the approach tractable for cosmological parameter estimation. We have shown that this extra shear-ratio information allows improved constraints on parameters which account for photometric redshift uncertainties, as well as cosmological parameters. Our approach here is an example of including theoretical uncertainties in the model, which is explored in detail by \citet{baldauf16}. Our case is an extreme one in that we allow complete freedom in $\Delta\Sigma$ below some scale. 

When it comes to using such methodology in an analysis of real data, there are several factors to consider that merit some discussion. Firstly, when using the point-mass marginalization, one must still choose a minimum scale $\rmin$ for which the $\xigm(r)$ prediction is trustworthy. We have discussed the rough scales on which typical modeling approaches are likely to break down at a level relevant to current and future large-scale structure surveys: $\sim 10-20\ \SI{}{\megaparsec}$ for a linear bias model, a few megaparsecs for a higher-order perturbative approach, while an HOD approach could potentially be reliable to tens or hundreds of kiloparsecs.

Ultimately, realistic, large volume galaxy simulations are required to inform this decision. Cosmological hydrodynamical simulations, which attempt to include some of the hydrodynmical processes important for galaxy formation, have advanced significantly in the past decade both in terms of simulation volume, and matching observed properties of the real universe (e.g. \citealt{schaye10,vogelsberger14,schaye15,springel18}).
However, there is still much uncertainty in the sub-grid prescriptions required to implement physical processes on scales below the resolution of these simulations. While uncertainty in the sub-grid modelling may not strongly impact the mass distribution on larger scales, it will impact the dependence of observable galaxy properties on that mass distribution, and hence the galaxy-galaxy lensing and clustering signals of a galaxy sample selected on observable properties.

It is likely therefore that empirical approaches where galaxies are added to gravity-only simulations using recipes calibrated against cosmological observables (e.g. \citealt{Tasitsiomi04,conroy06,hearin14,crocce15,derose19}) will continue to play an important role in understanding the relation between the distributions of galaxies and the distribution of matter in the Universe (see \citet{wechsler18} for a recent review). 

Such simulations are also likely required to estimate the impact of redshift evolution of lens properties across the width of lens redshift bins, a potential systematic effect when extracting shear-ration information using the methods presented here. We are hopeful however that given that galaxy-galaxy lensing analyses have typically been performed using lens galaxies with spectroscopic or high quality photometric redshifts, sufficiently narrow lens bins could usually be constructed.

Finally, we note the potential problems due to intrinsic galaxy alignments. If photometric redshift uncertainties in the source galaxy sample allow some overlap in redshift between the lens and source samples, there may be some net alignment of source galaxies' intrinsic shapes around lens galaxy positions \citep{hirata04,troxel14,Joachimi2015,blazek12}. Significant contamination from intrinsic alignments could bias the shear-ratio information extracted from small-scale galaxy-galaxy lensing signals, since the intrinsic alignment contribution will not scale according to \eqn{eq:shearratio}. So far, detections of this intrinsic alignment signal have largely been limited to bright, red galaxies (e.g. \citealt{mandelbaum06,hirata07,joachimi11,blazek11,singh16}). It is possible that this contamination can be mitigated by removing these galaxies from the source sample, through improved photo-z methods, or with modelling approaches (e.g. \citealt{crittenden01,hirata04,bridleking07,hui08,blazek15,blazek17}).

\section*{Acknowledgements}

Thanks to Gary Bernstein, Joe DeRose, Chris Hirata, Shivam Pandey, Judit Prat, Carles Sanchez and David Weinberg for useful discussions. JB is supported by a Swiss National Science Foundation Ambizione Fellowship. BJ is supported in part by the US Department of Energy grant desc0007901. This work used resources at the Ohio Supercomputing Center \citep{OhioSupercomputerCenter1987}.

\bibliographystyle{mnras}
\bibliography{main} 

\newcommand{\noop}[1]{}
\begin{thebibliography}{}
\makeatletter
\relax
\def\mn@urlcharsother{\let\do\@makeother \do\$\do\&\do\#\do\^\do\_\do\%\do\~}
\def\mn@doi{\begingroup\mn@urlcharsother \@ifnextchar [ {\mn@doi@}
  {\mn@doi@[]}}
\def\mn@doi@[#1]#2{\def\@tempa{#1}\ifx\@tempa\@empty \href
  {http://dx.doi.org/#2} {doi:#2}\else \href {http://dx.doi.org/#2} {#1}\fi
  \endgroup}
\def\mn@eprint#1#2{\mn@eprint@#1:#2::\@nil}
\def\mn@eprint@arXiv#1{\href {http://arxiv.org/abs/#1} {{\tt arXiv:#1}}}
\def\mn@eprint@dblp#1{\href {http://dblp.uni-trier.de/rec/bibtex/#1.xml}
  {dblp:#1}}
\def\mn@eprint@#1:#2:#3:#4\@nil{\def\@tempa {#1}\def\@tempb {#2}\def\@tempc
  {#3}\ifx \@tempc \@empty \let \@tempc \@tempb \let \@tempb \@tempa \fi \ifx
  \@tempb \@empty \def\@tempb {arXiv}\fi \@ifundefined
  {mn@eprint@\@tempb}{\@tempb:\@tempc}{\expandafter \expandafter \csname
  mn@eprint@\@tempb\endcsname \expandafter{\@tempc}}}

\bibitem[\protect\citeauthoryear{{Baldauf}, {Smith}, {Seljak}  \&
  {Mandelbaum}}{{Baldauf} et~al.}{2010}]{baldauf10}
{Baldauf} T.,  {Smith} R.~E.,  {Seljak} U.,   {Mandelbaum} R.,  2010, \mn@doi
  [\prd] {10.1103/PhysRevD.81.063531}, \href
  {http://adsabs.harvard.edu/abs/2010PhRvD..81f3531B} {81, 063531}

\bibitem[\protect\citeauthoryear{{Baldauf}, {Mirbabayi}, {Simonovi{\'c}}  \&
  {Zaldarriaga}}{{Baldauf} et~al.}{2016}]{baldauf16}
{Baldauf} T.,  {Mirbabayi} M.,  {Simonovi{\'c}} M.,   {Zaldarriaga} M.,  2016,
  arXiv e-prints, \href {http://adsabs.harvard.edu/abs/2016arXiv160200674B} {}

\bibitem[\protect\citeauthoryear{{Baltz}, {Marshall}  \& {Oguri}}{{Baltz}
  et~al.}{2009}]{baltz09}
{Baltz} E.~A.,  {Marshall} P.,   {Oguri} M.,  2009, \mn@doi [Journal of
  Cosmology and Astro-Particle Physics] {10.1088/1475-7516/2009/01/015}, \href
  {https://ui.adsabs.harvard.edu/\#abs/2009JCAP...01..015B} {2009, 015}

\bibitem[\protect\citeauthoryear{{Berlind} \& {Weinberg}}{{Berlind} \&
  {Weinberg}}{2002}]{berlind02}
{Berlind} A.~A.,  {Weinberg} D.~H.,  2002, \mn@doi [\apj] {10.1086/341469},
  \href {https://ui.adsabs.harvard.edu/\#abs/2002ApJ...575..587B} {575, 587}

\bibitem[\protect\citeauthoryear{{Bernstein} \& {Jain}}{{Bernstein} \&
  {Jain}}{2004}]{bernstein04}
{Bernstein} G.,  {Jain} B.,  2004, \mn@doi [\apj] {10.1086/379768}, \href
  {https://ui.adsabs.harvard.edu/\#abs/2004ApJ...600...17B} {600, 17}

\bibitem[\protect\citeauthoryear{{Blazek}, {McQuinn}  \& {Seljak}}{{Blazek}
  et~al.}{2011}]{blazek11}
{Blazek} J.,  {McQuinn} M.,   {Seljak} U.,  2011, \mn@doi [Journal of Cosmology
  and Astro-Particle Physics] {10.1088/1475-7516/2011/05/010}, \href
  {https://ui.adsabs.harvard.edu/\#abs/2011JCAP...05..010B} {2011, 010}

\bibitem[\protect\citeauthoryear{{Blazek}, {Mandelbaum}, {Seljak}  \&
  {Nakajima}}{{Blazek} et~al.}{2012}]{blazek12}
{Blazek} J.,  {Mandelbaum} R.,  {Seljak} U.,   {Nakajima} R.,  2012, \mn@doi
  [Journal of Cosmology and Astro-Particle Physics]
  {10.1088/1475-7516/2012/05/041}, \href
  {https://ui.adsabs.harvard.edu/\#abs/2012JCAP...05..041B} {2012, 041}

\bibitem[\protect\citeauthoryear{{Blazek}, {Vlah}  \& {Seljak}}{{Blazek}
  et~al.}{2015}]{blazek15}
{Blazek} J.,  {Vlah} Z.,   {Seljak} U.,  2015, \mn@doi [\jcap]
  {10.1088/1475-7516/2015/08/015}, \href
  {http://adsabs.harvard.edu/abs/2015JCAP...08..015B} {8, 015}

\bibitem[\protect\citeauthoryear{{Blazek}, {MacCrann}, {Troxel}  \&
  {Fang}}{{Blazek} et~al.}{2017}]{blazek17}
{Blazek} J.,  {MacCrann} N.,  {Troxel} M.~A.,   {Fang} X.,  2017, arXiv
  e-prints, \href {https://ui.adsabs.harvard.edu/\#abs/2017arXiv170809247B} {p.
  arXiv:1708.09247}

\bibitem[\protect\citeauthoryear{{Brainerd}, {Blandford}  \&
  {Smail}}{{Brainerd} et~al.}{1996}]{brainerd96}
{Brainerd} T.~G.,  {Blandford} R.~D.,   {Smail} I.,  1996, \mn@doi [\apj]
  {10.1086/177537}, \href {http://adsabs.harvard.edu/abs/1996ApJ...466..623B}
  {466, 623}

\bibitem[\protect\citeauthoryear{{Bridle} \& {King}}{{Bridle} \&
  {King}}{2007}]{bridleking07}
{Bridle} S.,  {King} L.,  2007, \mn@doi [New Journal of Physics]
  {10.1088/1367-2630/9/12/444}, \href
  {https://ui.adsabs.harvard.edu/\#abs/2007NJPh....9..444B} {9, 444}

\bibitem[\protect\citeauthoryear{{Bridle}, {Crittenden}, {Melchiorri},
  {Hobson}, {Kneissl}  \& {Lasenby}}{{Bridle} et~al.}{2002}]{bridle02}
{Bridle} S.~L.,  {Crittenden} R.,  {Melchiorri} A.,  {Hobson} M.~P.,  {Kneissl}
  R.,   {Lasenby} A.~N.,  2002, \mn@doi [\mnras]
  {10.1046/j.1365-8711.2002.05709.x}, \href
  {https://ui.adsabs.harvard.edu/\#abs/2002MNRAS.335.1193B} {335, 1193}

\bibitem[\protect\citeauthoryear{{Bridle} et~al.,}{{Bridle}
  et~al.}{2009}]{great08handbook}
{Bridle} S.,  et~al., 2009, \mn@doi [Annals of Applied Statistics]
  {10.1214/08-AOAS222}, \href
  {http://adsabs.harvard.edu/abs/2009AnApS...3....6B} {3, 6}

\bibitem[\protect\citeauthoryear{{Cacciato}, {van den Bosch}, {More}, {Mo}  \&
  {Yang}}{{Cacciato} et~al.}{2013}]{cacciato13}
{Cacciato} M.,  {van den Bosch} F.~C.,  {More} S.,  {Mo} H.,   {Yang} X.,
  2013, \mn@doi [\mnras] {10.1093/mnras/sts525}, \href
  {https://ui.adsabs.harvard.edu/\#abs/2013MNRAS.430..767C} {430, 767}

\bibitem[\protect\citeauthoryear{{Choi}, {Tyson}, {Morrison}, {Jee}, {Schmidt},
  {Margoniner}  \& {Wittman}}{{Choi} et~al.}{2012}]{choi2012}
{Choi} A.,  {Tyson} J.~A.,  {Morrison} C.~B.,  {Jee} M.~J.,  {Schmidt} S.~J.,
  {Margoniner} V.~E.,   {Wittman} D.~M.,  2012, \mn@doi [\apj]
  {10.1088/0004-637X/759/2/101}, \href
  {https://ui.adsabs.harvard.edu/\#abs/2012ApJ...759..101C} {759, 101}

\bibitem[\protect\citeauthoryear{{Clampitt} et~al.,}{{Clampitt}
  et~al.}{2017}]{clampitt17}
{Clampitt} J.,  et~al., 2017, \mn@doi [\mnras] {10.1093/mnras/stw2988}, \href
  {https://ui.adsabs.harvard.edu/\#abs/2017MNRAS.465.4204C} {465, 4204}

\bibitem[\protect\citeauthoryear{{Conroy}, {Wechsler}  \& {Kravtsov}}{{Conroy}
  et~al.}{2006}]{conroy06}
{Conroy} C.,  {Wechsler} R.~H.,   {Kravtsov} A.~V.,  2006, \mn@doi [\apj]
  {10.1086/503602}, \href
  {https://ui.adsabs.harvard.edu/\#abs/2006ApJ...647..201C} {647, 201}

\bibitem[\protect\citeauthoryear{{Crittenden}, {Natarajan}, {Pen}  \&
  {Theuns}}{{Crittenden} et~al.}{2001}]{crittenden01}
{Crittenden} R.~G.,  {Natarajan} P.,  {Pen} U.-L.,   {Theuns} T.,  2001,
  \mn@doi [\apj] {10.1086/322370}, \href
  {https://ui.adsabs.harvard.edu/\#abs/2001ApJ...559..552C} {559, 552}

\bibitem[\protect\citeauthoryear{{Crocce}, {Castander}, {Gazta{\~n}aga},
  {Fosalba}  \& {Carretero}}{{Crocce} et~al.}{2015}]{crocce15}
{Crocce} M.,  {Castander} F.~J.,  {Gazta{\~n}aga} E.,  {Fosalba} P.,
  {Carretero} J.,  2015, \mn@doi [\mnras] {10.1093/mnras/stv1708}, \href
  {https://ui.adsabs.harvard.edu/\#abs/2015MNRAS.453.1513C} {453, 1513}

\bibitem[\protect\citeauthoryear{{DES Collaboration} et~al.}{{DES
  Collaboration} et~al.}{2017}]{keypaper}
{DES Collaboration} et~al., 2017, to be submitted to Phys. Rev. D

\bibitem[\protect\citeauthoryear{{DeRose} et~al.,}{{DeRose}
  et~al.}{2019}]{derose19}
{DeRose} J.,  et~al., 2019, arXiv e-prints, \href
  {https://ui.adsabs.harvard.edu/\#abs/2019arXiv190102401D} {p.
  arXiv:1901.02401}

\bibitem[\protect\citeauthoryear{{Desjacques}, {Jeong}  \&
  {Schmidt}}{{Desjacques} et~al.}{2018}]{desjacques18}
{Desjacques} V.,  {Jeong} D.,   {Schmidt} F.,  2018, \mn@doi [\physrep]
  {10.1016/j.physrep.2017.12.002}, \href
  {https://ui.adsabs.harvard.edu/\#abs/2018PhR...733....1D} {733, 1}

\bibitem[\protect\citeauthoryear{{Duffy}, {Schaye}, {Kay}  \& {Dalla
  Vecchia}}{{Duffy} et~al.}{2008}]{duffy08}
{Duffy} A.~R.,  {Schaye} J.,  {Kay} S.~T.,   {Dalla Vecchia} C.,  2008, \mn@doi
  [\mnras] {10.1111/j.1745-3933.2008.00537.x}, \href
  {https://ui.adsabs.harvard.edu/\#abs/2008MNRAS.390L..64D} {390, L64}

\bibitem[\protect\citeauthoryear{{Fry} \& {Gaztanaga}}{{Fry} \&
  {Gaztanaga}}{1993}]{fry93}
{Fry} J.~N.,  {Gaztanaga} E.,  1993, \mn@doi [\apj] {10.1086/173015}, \href
  {http://adsabs.harvard.edu/abs/1993ApJ...413..447F} {413, 447}

\bibitem[\protect\citeauthoryear{{Hearin}, {Watson}, {Becker}, {Reyes},
  {Berlind}  \& {Zentner}}{{Hearin} et~al.}{2014}]{hearin14}
{Hearin} A.~P.,  {Watson} D.~F.,  {Becker} M.~R.,  {Reyes} R.,  {Berlind}
  A.~A.,   {Zentner} A.~R.,  2014, \mn@doi [\mnras] {10.1093/mnras/stu1443},
  \href {https://ui.adsabs.harvard.edu/\#abs/2014MNRAS.444..729H} {444, 729}

\bibitem[\protect\citeauthoryear{{Heymans} et~al.,}{{Heymans}
  et~al.}{2012}]{heymans12}
{Heymans} C.,  et~al., 2012, \mn@doi [\mnras]
  {10.1111/j.1365-2966.2012.21952.x}, \href
  {http://adsabs.harvard.edu/abs/2012MNRAS.427..146H} {427, 146}

\bibitem[\protect\citeauthoryear{{Hildebrandt} et~al.,}{{Hildebrandt}
  et~al.}{2017}]{hildebrandt17}
{Hildebrandt} H.,  et~al., 2017, \mn@doi [\mnras] {10.1093/mnras/stw2805},
  \href {http://adsabs.harvard.edu/abs/2017MNRAS.465.1454H} {465, 1454}

\bibitem[\protect\citeauthoryear{{Hirata} et~al.,}{{Hirata}
  et~al.}{2004}]{hirata04}
{Hirata} C.~M.,  et~al., 2004, \mn@doi [\mnras]
  {10.1111/j.1365-2966.2004.08090.x}, \href
  {http://adsabs.harvard.edu/abs/2004MNRAS.353..529H} {353, 529}

\bibitem[\protect\citeauthoryear{{Hirata}, {Mandelbaum}, {Ishak}, {Seljak},
  {Nichol}, {Pimbblet}, {Ross}  \& {Wake}}{{Hirata} et~al.}{2007}]{hirata07}
{Hirata} C.~M.,  {Mandelbaum} R.,  {Ishak} M.,  {Seljak} U.,  {Nichol} R.,
  {Pimbblet} K.~A.,  {Ross} N.~P.,   {Wake} D.,  2007, \mn@doi [\mnras]
  {10.1111/j.1365-2966.2007.12312.x}, \href
  {https://ui.adsabs.harvard.edu/\#abs/2007MNRAS.381.1197H} {381, 1197}

\bibitem[\protect\citeauthoryear{{Hoyle} et~al.,}{{Hoyle}
  et~al.}{2018}]{y1photoz}
{Hoyle} B.,  et~al., 2018, \mn@doi [\mnras] {10.1093/mnras/sty957}, \href
  {https://ui.adsabs.harvard.edu/\#abs/2018MNRAS.478..592H} {478, 592}

\bibitem[\protect\citeauthoryear{{Hu} \& {Jain}}{{Hu} \& {Jain}}{2004}]{hu04}
{Hu} W.,  {Jain} B.,  2004, \mn@doi [\prd] {10.1103/PhysRevD.70.043009}, \href
  {http://adsabs.harvard.edu/abs/2004PhRvD..70d3009H} {70, 043009}

\bibitem[\protect\citeauthoryear{{Hui} \& {Zhang}}{{Hui} \&
  {Zhang}}{2008}]{hui08}
{Hui} L.,  {Zhang} J.,  2008, \mn@doi [\apj] {10.1086/589872}, \href
  {http://adsabs.harvard.edu/abs/2008ApJ...688..742H} {688, 742}

\bibitem[\protect\citeauthoryear{{Jain} \& {Taylor}}{{Jain} \&
  {Taylor}}{2003}]{jain03}
{Jain} B.,  {Taylor} A.,  2003, \mn@doi [\prl] {10.1103/PhysRevLett.91.141302},
  \href {https://ui.adsabs.harvard.edu/\#abs/2003PhRvL..91n1302J} {91, 141302}

\bibitem[\protect\citeauthoryear{{Joachimi}, {Mandelbaum}, {Abdalla}  \&
  {Bridle}}{{Joachimi} et~al.}{2011}]{joachimi11}
{Joachimi} B.,  {Mandelbaum} R.,  {Abdalla} F.~B.,   {Bridle} S.~L.,  2011,
  \mn@doi [\aap] {10.1051/0004-6361/201015621}, 527, A26

\bibitem[\protect\citeauthoryear{{Joachimi} et~al.,}{{Joachimi}
  et~al.}{2015}]{Joachimi2015}
{Joachimi} B.,  et~al., 2015, preprint, \href
  {http://adsabs.harvard.edu/abs/2015arXiv150405456J} {} (\mn@eprint {arXiv}
  {1504.05456})

\bibitem[\protect\citeauthoryear{{Joudaki} et~al.,}{{Joudaki}
  et~al.}{2018}]{joudaki18}
{Joudaki} S.,  et~al., 2018, \mn@doi [\mnras] {10.1093/mnras/stx2820}, \href
  {https://ui.adsabs.harvard.edu/\#abs/2018MNRAS.474.4894J} {474, 4894}

\bibitem[\protect\citeauthoryear{{Kaiser}}{{Kaiser}}{1984}]{kaiser84}
{Kaiser} N.,  1984, \mn@doi [\apjl] {10.1086/184341}, \href
  {http://adsabs.harvard.edu/abs/1984ApJ...284L...9K} {284, L9}

\bibitem[\protect\citeauthoryear{{Kitching} et~al.,}{{Kitching}
  et~al.}{2012}]{great10}
{Kitching} T.~D.,  et~al., 2012, \mn@doi [\mnras]
  {10.1111/j.1365-2966.2012.21095.x}, \href
  {http://adsabs.harvard.edu/abs/2012MNRAS.423.3163K} {423, 3163}

\bibitem[\protect\citeauthoryear{{Krause} et~al.,}{{Krause}
  et~al.}{2017}]{methodpaper}
{Krause} E.,  et~al., 2017, arXiv e-prints, \href
  {https://ui.adsabs.harvard.edu/\#abs/2017arXiv170609359K} {p.
  arXiv:1706.09359}

\bibitem[\protect\citeauthoryear{{Leauthaud} et~al.,}{{Leauthaud}
  et~al.}{2012}]{leauthaud12}
{Leauthaud} A.,  et~al., 2012, \mn@doi [\apj] {10.1088/0004-637X/744/2/159},
  \href {https://ui.adsabs.harvard.edu/\#abs/2012ApJ...744..159L} {744, 159}

\bibitem[\protect\citeauthoryear{{Mandelbaum}}{{Mandelbaum}}{2018}]{mandelbaum18}
{Mandelbaum} R.,  2018, \mn@doi [Annual Review of Astronomy and Astrophysics]
  {10.1146/annurev-astro-081817-051928}, \href
  {https://ui.adsabs.harvard.edu/\#abs/2018ARA&A..56..393M} {56, 393}

\bibitem[\protect\citeauthoryear{{Mandelbaum}, {Hirata}, {Ishak}, {Seljak}  \&
  {Brinkmann}}{{Mandelbaum} et~al.}{2006}]{mandelbaum06}
{Mandelbaum} R.,  {Hirata} C.~M.,  {Ishak} M.,  {Seljak} U.,   {Brinkmann} J.,
  2006, \mn@doi [\mnras] {10.1111/j.1365-2966.2005.09946.x}, \href
  {https://ui.adsabs.harvard.edu/\#abs/2006MNRAS.367..611M} {367, 611}

\bibitem[\protect\citeauthoryear{{Mandelbaum}, {Slosar}, {Baldauf}, {Seljak},
  {Hirata}, {Nakajima}, {Reyes}  \& {Smith}}{{Mandelbaum}
  et~al.}{2013}]{mandelbaum13}
{Mandelbaum} R.,  {Slosar} A.,  {Baldauf} T.,  {Seljak} U.,  {Hirata} C.~M.,
  {Nakajima} R.,  {Reyes} R.,   {Smith} R.~E.,  2013, \mn@doi [\mnras]
  {10.1093/mnras/stt572}, \href
  {http://adsabs.harvard.edu/abs/2013MNRAS.432.1544M} {432, 1544}

\bibitem[\protect\citeauthoryear{{Mandelbaum} et~al.,}{{Mandelbaum}
  et~al.}{2014}]{great3handbook}
{Mandelbaum} R.,  et~al., 2014, \mn@doi [\apjs] {10.1088/0067-0049/212/1/5},
  \href {http://adsabs.harvard.edu/abs/2014ApJS..212....5M} {212, 5}

\bibitem[\protect\citeauthoryear{{Mandelbaum} et~al.,}{{Mandelbaum}
  et~al.}{2018}]{mandelbaum18b}
{Mandelbaum} R.,  et~al., 2018, \mn@doi [\pasj] {10.1093/pasj/psx130}, \href
  {http://ads.nao.ac.jp/abs/2018PASJ...70S..25M} {70, S25}

\bibitem[\protect\citeauthoryear{{Miyatake}, {Madhavacheril}, {Sehgal},
  {Slosar}, {Spergel}, {Sherwin}  \& {van Engelen}}{{Miyatake}
  et~al.}{2017}]{miyatake17}
{Miyatake} H.,  {Madhavacheril} M.~S.,  {Sehgal} N.,  {Slosar} A.,  {Spergel}
  D.~N.,  {Sherwin} B.,   {van Engelen} A.,  2017, \mn@doi [\prl]
  {10.1103/PhysRevLett.118.161301}, \href
  {https://ui.adsabs.harvard.edu/\#abs/2017PhRvL.118p1301M} {118, 161301}

\bibitem[\protect\citeauthoryear{{Nishimichi} et~al.,}{{Nishimichi}
  et~al.}{2018}]{nishmichi18}
{Nishimichi} T.,  et~al., 2018, arXiv e-prints, \href
  {https://ui.adsabs.harvard.edu/\#abs/2018arXiv181109504N} {p.
  arXiv:1811.09504}

\bibitem[\protect\citeauthoryear{{Ohio Supercomputer Center}}{{Ohio
  Supercomputer Center}}{1987}]{OhioSupercomputerCenter1987}
{Ohio Supercomputer Center} 1987, Ohio Supercomputer Center,
  \url{http://osc.edu/ark:/19495/f5s1ph73}

\bibitem[\protect\citeauthoryear{{Peacock} \& {Smith}}{{Peacock} \&
  {Smith}}{2000}]{peacock00}
{Peacock} J.~A.,  {Smith} R.~E.,  2000, \mn@doi [\mnras]
  {10.1046/j.1365-8711.2000.03779.x}, \href
  {http://adsabs.harvard.edu/abs/2000MNRAS.318.1144P} {318, 1144}

\bibitem[\protect\citeauthoryear{{Prat} et~al.,}{{Prat} et~al.}{2018a}]{prat18}
{Prat} J.,  et~al., 2018a, arXiv e-prints, \href
  {https://ui.adsabs.harvard.edu/\#abs/2018arXiv181002212P} {p.
  arXiv:1810.02212}

\bibitem[\protect\citeauthoryear{{Prat} et~al.,}{{Prat}
  et~al.}{2018b}]{pratsanchez18}
{Prat} J.,  et~al., 2018b, \mn@doi [\prd] {10.1103/PhysRevD.98.042005}, \href
  {https://ui.adsabs.harvard.edu/\#abs/2018PhRvD..98d2005P} {98, 042005}

\bibitem[\protect\citeauthoryear{{Schaye} et~al.,}{{Schaye}
  et~al.}{2010}]{schaye10}
{Schaye} J.,  et~al., 2010, \mn@doi [\mnras]
  {10.1111/j.1365-2966.2009.16029.x}, \href
  {http://adsabs.harvard.edu/abs/2010MNRAS.402.1536S} {402, 1536}

\bibitem[\protect\citeauthoryear{{Schaye} et~al.,}{{Schaye}
  et~al.}{2015}]{schaye15}
{Schaye} J.,  et~al., 2015, \mn@doi [\mnras] {10.1093/mnras/stu2058}, \href
  {http://adsabs.harvard.edu/abs/2015MNRAS.446..521S} {446, 521}

\bibitem[\protect\citeauthoryear{{Seljak}}{{Seljak}}{2000}]{seljak00}
{Seljak} U.,  2000, \mn@doi [\mnras] {10.1046/j.1365-8711.2000.03715.x}, \href
  {http://adsabs.harvard.edu/abs/2000MNRAS.318..203S} {318, 203}

\bibitem[\protect\citeauthoryear{{Singh} \& {Mandelbaum}}{{Singh} \&
  {Mandelbaum}}{2016}]{singh16}
{Singh} S.,  {Mandelbaum} R.,  2016, \mn@doi [\mnras] {10.1093/mnras/stw144},
  \href {https://ui.adsabs.harvard.edu/\#abs/2016MNRAS.457.2301S} {457, 2301}

\bibitem[\protect\citeauthoryear{{Singh}, {Mandelbaum}, {Seljak},
  {Rodr{\'\i}guez-Torres}  \& {Slosar}}{{Singh} et~al.}{2018}]{singh18}
{Singh} S.,  {Mandelbaum} R.,  {Seljak} U.,  {Rodr{\'\i}guez-Torres} S.,
  {Slosar} A.,  2018, arXiv e-prints, \href
  {https://ui.adsabs.harvard.edu/\#abs/2018arXiv181106499S} {p.
  arXiv:1811.06499}

\bibitem[\protect\citeauthoryear{{Smith} et~al.,}{{Smith}
  et~al.}{2003}]{smith03}
{Smith} R.~E.,  et~al., 2003, \mn@doi [\mnras]
  {10.1046/j.1365-8711.2003.06503.x}, \href
  {http://adsabs.harvard.edu/abs/2003MNRAS.341.1311S} {341, 1311}

\bibitem[\protect\citeauthoryear{{Smith}, {Scoccimarro}  \& {Sheth}}{{Smith}
  et~al.}{2007}]{smith07}
{Smith} R.~E.,  {Scoccimarro} R.,   {Sheth} R.~K.,  2007, \mn@doi [\prd]
  {10.1103/PhysRevD.75.063512}, \href
  {https://ui.adsabs.harvard.edu/\#abs/2007PhRvD..75f3512S} {75, 063512}

\bibitem[\protect\citeauthoryear{{Springel} et~al.,}{{Springel}
  et~al.}{2018}]{springel18}
{Springel} V.,  et~al., 2018, \mn@doi [\mnras] {10.1093/mnras/stx3304}, \href
  {http://adsabs.harvard.edu/abs/2018MNRAS.475..676S} {475, 676}

\bibitem[\protect\citeauthoryear{{Takahashi}, {Sato}, {Nishimichi}, {Taruya}
  \& {Oguri}}{{Takahashi} et~al.}{2012}]{takahashi2012}
{Takahashi} R.,  {Sato} M.,  {Nishimichi} T.,  {Taruya} A.,   {Oguri} M.,
  2012, \mn@doi [\apj] {10.1088/0004-637X/761/2/152}, \href
  {http://adsabs.harvard.edu/abs/2012ApJ...761..152T} {761, 152}

\bibitem[\protect\citeauthoryear{{Tanaka} et~al.,}{{Tanaka}
  et~al.}{2018}]{tanaka18}
{Tanaka} M.,  et~al., 2018, \mn@doi [\pasj] {10.1093/pasj/psx077}, \href
  {http://ads.nao.ac.jp/abs/2018PASJ...70S...9T} {70, S9}

\bibitem[\protect\citeauthoryear{{Tasitsiomi}, {Kravtsov}, {Wechsler}  \&
  {Primack}}{{Tasitsiomi} et~al.}{2004}]{Tasitsiomi04}
{Tasitsiomi} A.,  {Kravtsov} A.~V.,  {Wechsler} R.~H.,   {Primack} J.~R.,
  2004, \mn@doi [\apj] {10.1086/423784}, \href
  {http://adsabs.harvard.edu/abs/2004ApJ...614..533T} {614, 533}

\bibitem[\protect\citeauthoryear{{Taylor}, {Kitching}, {Bacon}  \&
  {Heavens}}{{Taylor} et~al.}{2007}]{taylor07}
{Taylor} A.~N.,  {Kitching} T.~D.,  {Bacon} D.~J.,   {Heavens} A.~F.,  2007,
  \mn@doi [\mnras] {10.1111/j.1365-2966.2006.11257.x}, \href
  {https://ui.adsabs.harvard.edu/\#abs/2007MNRAS.374.1377T} {374, 1377}

\bibitem[\protect\citeauthoryear{{Tinker}, {Robertson}, {Kravtsov}, {Klypin},
  {Warren}, {Yepes}  \& {Gottl{\"o}ber}}{{Tinker} et~al.}{2010}]{tinker10}
{Tinker} J.~L.,  {Robertson} B.~E.,  {Kravtsov} A.~V.,  {Klypin} A.,  {Warren}
  M.~S.,  {Yepes} G.,   {Gottl{\"o}ber} S.,  2010, \mn@doi [\apj]
  {10.1088/0004-637X/724/2/878}, \href
  {http://adsabs.harvard.edu/abs/2010ApJ...724..878T} {724, 878}

\bibitem[\protect\citeauthoryear{{Troxel} \& {Ishak}}{{Troxel} \&
  {Ishak}}{2014}]{troxel14}
{Troxel} M.~A.,  {Ishak} M.,  2014, \mn@doi [\prd]
  {10.1103/PhysRevD.89.063528}, \href
  {http://adsabs.harvard.edu/abs/2014PhRvD..89f3528T} {89, 063528}

\bibitem[\protect\citeauthoryear{{Tyson}, {Valdes}, {Jarvis}  \&
  {Mills}}{{Tyson} et~al.}{1984}]{tyson84}
{Tyson} J.~A.,  {Valdes} F.,  {Jarvis} J.~F.,   {Mills} Jr. A.~P.,  1984,
  \mn@doi [\apjl] {10.1086/184285}, \href
  {http://adsabs.harvard.edu/abs/1984ApJ...281L..59T} {281, L59}

\bibitem[\protect\citeauthoryear{{Velander} et~al.,}{{Velander}
  et~al.}{2014}]{velander14}
{Velander} M.,  et~al., 2014, \mn@doi [\mnras] {10.1093/mnras/stt2013}, \href
  {https://ui.adsabs.harvard.edu/\#abs/2014MNRAS.437.2111V} {437, 2111}

\bibitem[\protect\citeauthoryear{{Viola} et~al.,}{{Viola}
  et~al.}{2015}]{viola15}
{Viola} M.,  et~al., 2015, \mn@doi [\mnras] {10.1093/mnras/stv1447}, \href
  {https://ui.adsabs.harvard.edu/\#abs/2015MNRAS.452.3529V} {452, 3529}

\bibitem[\protect\citeauthoryear{{Vogelsberger} et~al.,}{{Vogelsberger}
  et~al.}{2014}]{vogelsberger14}
{Vogelsberger} M.,  et~al., 2014, \mn@doi [\mnras] {10.1093/mnras/stu1536},
  \href {http://adsabs.harvard.edu/abs/2014MNRAS.444.1518V} {444, 1518}

\bibitem[\protect\citeauthoryear{{Wechsler} \& {Tinker}}{{Wechsler} \&
  {Tinker}}{2018}]{wechsler18}
{Wechsler} R.~H.,  {Tinker} J.~L.,  2018, \mn@doi [Annual Review of Astronomy
  and Astrophysics] {10.1146/annurev-astro-081817-051756}, \href
  {https://ui.adsabs.harvard.edu/\#abs/2018ARA&A..56..435W} {56, 435}

\bibitem[\protect\citeauthoryear{{Weinberg}, {Mortonson}, {Eisenstein},
  {Hirata}, {Riess}  \& {Rozo}}{{Weinberg} et~al.}{2013}]{weinberg13}
{Weinberg} D.~H.,  {Mortonson} M.~J.,  {Eisenstein} D.~J.,  {Hirata} C.,
  {Riess} A.~G.,   {Rozo} E.,  2013, \mn@doi [\physrep]
  {10.1016/j.physrep.2013.05.001}, \href
  {http://adsabs.harvard.edu/abs/2013PhR...530...87W} {530, 87}

\bibitem[\protect\citeauthoryear{{Wibking} et~al.,}{{Wibking}
  et~al.}{2019}]{wibking19}
{Wibking} B.~D.,  et~al., 2019, \mn@doi [\mnras] {10.1093/mnras/sty2258}, \href
  {https://ui.adsabs.harvard.edu/\#abs/2019MNRAS.484..989W} {484, 989}

\bibitem[\protect\citeauthoryear{Zuntz et~al.,}{Zuntz et~al.}{2015}]{cosmosis}
Zuntz J.,  et~al., 2015, \mn@doi [Astronomy and Computing]
  {http://dx.doi.org/10.1016/j.ascom.2015.05.005}, 12, 45

\bibitem[\protect\citeauthoryear{{van Uitert} et~al.,}{{van Uitert}
  et~al.}{2016}]{vanuitert16}
{van Uitert} E.,  et~al., 2016, \mn@doi [\mnras] {10.1093/mnras/stw747}, \href
  {https://ui.adsabs.harvard.edu/\#abs/2016MNRAS.459.3251V} {459, 3251}

\bibitem[\protect\citeauthoryear{{van Uitert} et~al.,}{{van Uitert}
  et~al.}{2018}]{vanuitert18}
{van Uitert} E.,  et~al., 2018, \mn@doi [\mnras] {10.1093/mnras/sty551}, \href
  {https://ui.adsabs.harvard.edu/\#abs/2018MNRAS.476.4662V} {476, 4662}

\makeatother
\end{thebibliography}

\bsp
\label{lastpage}
\end{document}